# Twisted light affects ultrafast demagnetization

**Authors:** Eva Prinz[1], Paul Herrgen[1*], Jonas Hoefer[1], Benjamin Stadtmüller[1,2], Martin Aeschlimann[1]

**Affiliations:**

[1]Department of Physics and Research Center OPTIMAS, RPTU University Kaiserslautern-Landau, 67663 Kaiserslautern, Germany

[2]Experimentalphysik II, Institute of Physics, Augsburg University, D-86135 Augsburg, Germany.

*Corresponding author. Email: pherrgen@rptu.de

**Irradiation with an ultrashort laser pulse can completely destroy the magnetic order of ferromagnetic thin films on the femtosecond timescale. This phenomenon holds great potential for ultrafast spintronics and information processing and is an active field of research. It is still an open question if the angular momentum of light can support this effect. While it has been shown that the spin of light only has a negligible influence, we experimentally demonstrate the influence of photonic orbital angular momentum (OAM) on the magnetization dynamics of a thin nickel film. Our results reveal that the photonic OAM affects the demagnetization behavior by a non-coherent, secondary process within the first hundreds of femtoseconds depending on the handedness of the OAM with respect to the direction of the sample magnetization.**

## Introduction

Photons can carry two types of angular momentum. As bosons, they always have a spin $S = s\hbar = \pm 1$ (with the spin quantum number $s\hbar = \pm 1$ and the reduced Planck constant $\hbar$) that is associated with left- or right-handed circular polarization, respectively. In addition, they can also carry an orbital angular momentum (OAM)[1-3] $L = l\hbar$ (with the OAM quantum number $l \in \mathbb{Z}$) that leads to characteristic annular beam profiles with a singularity at the center and a spiral energy flow. It has been shown that photonic OAM, often referred to as twisted light, can be transferred to matter to mechanically rotate micro-particles[4] as well as to influence higher-order electronic transitions[5]. Both types of angular momentum are used for various applications, such as data transmission[6-9], quantum information technology[10-12], and micromanipulations in optical tweezers[13-15].

Angular momentum is directly related to magnetic moments, the source of magnetism. Thus, the vision to control magnetization and its optically induced dynamics[16] with photonic OAM is very intuitive. In general, the ultrafast, femtosecond time scale magnetization dynamics is triggered by irradiation of magnetic samples with ultrashort laser pulses. Here, energy can be deposited within tens of femtoseconds onto the material while the conservation of angular momentum, which must be transferred from the electron spin to the lattice[17], is a limiting factor for the speed of the demagnetization. During this process, photonic angular momentum from the laser pulse could affect the angular momentum balance in the material. First attempts explored the influence of the light's spin on the magnetization dynamics. The spin of light, however, has been shown to have no measurable influence on the ultrafast demagnetization of simple ferromagnetic thin films such as Ni[18,19], but it is crucial for the so-called helicity-dependent all-optical switching (HD-AOS)[20,21]. This phenomenon has been observed in ferromagnetic rare-earth transition-metal alloys as well as in ferromagnetic multilayer systems and is solely based on the spin degree of freedom of the photons, either through the inverse Faraday effect (IFE)[22,23] or through magnetic circular dichroism (MCD)[23-25].

In our work, we aim to gain new insights into the pressing question of whether photonic OAM can also be used to control magnetization on ultrafast timescales. Recent publications have already demonstrated a dichroism effect for terahertz radiation with OAM in the rare earth iron garnet $Dy_3Fe_5O_{12}$[26] and the possibility of modifying the magnetic anisotropy of an interlayer exchange coupling system with OAM light[27]. In our study, we focus directly on the transfer of photonic OAM into magnetic materials at the time scale of the optical excitation. First, we illuminated a thin nickel film with linearly polarized ultrashort laser pulses carrying varying orders of OAM. By recording the magnetic response of the sample, we observed that OAM either supports or obstructs the demagnetization process, depending on the direction of the OAM vector relative to the sample magnetization. Crucially, we observed no measurable impact on the magnetization dynamics when repeating the same experiment on a thin permalloy (FeNi) film. This finding provides evidence for the reliability of our experimental procedure and yields insights into possible OAM conversion processes from photons to the spin system of ferromagnets.

## Results

### Magnetic and nonmagnetic signals in time-resolved MOKE experiments with OAM pump pulses

To investigate the influence of photonic OAM on ultrafast laser-induced demagnetization, we chose the longitudinal time-resolved magneto-optical Kerr effect (TR-MOKE) technique. The sample consists of a 10 nm thick polycrystalline nickel film deposited on a MgO substrate and is protected by a 100 nm $SiO_2$ capping layer. The sample was optically excited with ultrashort ($\tau_{\text{pump}} \leq 35$ fs), linearly polarized laser pulses from a Ti:Sa amplifier with a central wavelength of $\lambda_{\text{pump}} = 800$ nm carrying an OAM of $l \in [-8, 8]$. We probed the magnetic response of a position on the ring of the OAM mode (see Fig. 3 and Supplementary Figs. 3-5 for beam camera images) with the linearly polarized second harmonic of the amplifier with a central wavelength of $\lambda_{\text{probe}} = 400$ nm ($\tau_{\text{probe}} \leq 75$ fs) that did not carry an OAM ($l_{\text{probe}} = 0$).

To unambiguously obtain the intrinsic magnetic response in TR-MOKE experiments, it is essential to disentangle the magnetic from nonmagnetic contributions in the magnetooptical signal. Therefore, it is common practice to perform the time-resolved Kerr experiment for opposite magnetization directions ($M^+$, $M^-$) of the ferromagnetic sample. The pure magnetic material response is then extracted by calculating the difference of the two experimental time traces ($M^+ - M^-$). This procedure eliminates any signal contributions that are independent of the direction of the sample magnetization. On the other hand, the manifold of nonmagnetic contributions can be obtained by calculating the sum of the two curves ($M^+ + M^-$). The most commonly observed nonmagnetic contribution is the so-called dichroic bleaching or state-blocking effect[18,28,29], where the pump pulse saturates optical transitions, thus making them unavailable to the probe pulse. This phenomenon is especially relevant for monochromatic experiments where the pump and probe pulses share the same wavelength. In experiments with circularly polarized pump pulses, additional effects such as the specular inverse Faraday effect (SIFE) and specular optical Kerr effect (SOKE)[30] lead to a peak around zero pulse delay ($t_0$). In a simplified picture, the effects refer to a transfer of angular momentum from circularly polarized photons to the electron's OAM, which influences the experimental signal[19]. In addition, when the linear polarization directions of the pump and probe beams are not perfectly parallel, an electro-optical Kerr effect induced in the sample is likely to affect the measurement signal. For strong optical excitations, another nonmagnetic effect also needs to be considered, namely the pump-induced birefringence of the air in the region where pump and probe beam overlap in front of the sample[31-34]. The high-intensity pump pulse's electric field interacts with the nonlinear air molecules (i.e., $N_2$, $O_2$, and $CO_2$)[34], which experience a torque. The molecules align with the propagation direction of the pump pulse and the probe pulse is affected by the birefringence of the air. This effect occurs even for parallel linear pump and probe polarizations.

In addition to these conventional magnetic and nonmagnetic contributions, our experiment also needs to take into account the direction of the OAM vector $\boldsymbol{L}$ with respect to the sample magnetization $\boldsymbol{M}$. This leads to a total of four different measurement geometries. For two of them, the projection of the OAM vector onto the magnetization direction of the sample and the magnetization are parallel ($\boldsymbol{L} \cdot \boldsymbol{M} > 0$, see Fig. 1**a**, **b** (① & ②)) and the light's OAM *supports* the demagnetization process. For the other two geometries, the directions are antiparallel ($\boldsymbol{L} \cdot \boldsymbol{M} < 0$, Fig. 1**c**, **d** (③ & ④)), and the OAM *obstructs* the demagnetization. To reveal this influence of the photonic OAM, the differences of the supporting and obstructing cases must be considered, respectively (see methods for details).

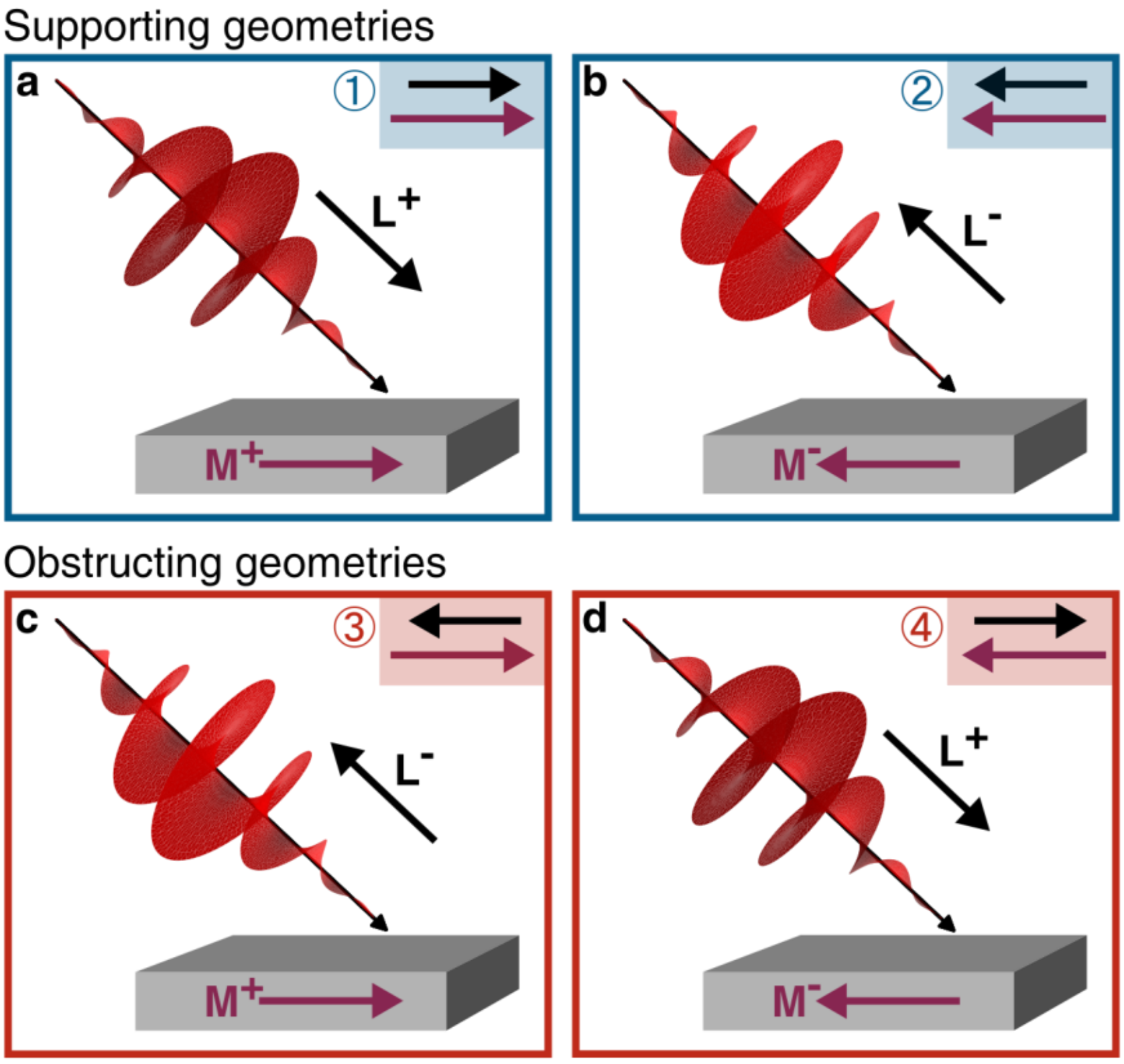

**Fig. 1: Schematic of the four different measurement geometries used to investigate the influence of optical OAM on laser-induced demagnetization.** **a**, **b**, Supporting geometries (① & ②). In these geometries, the direction of the sample magnetization $\boldsymbol{M}$ and the projection of the OAM vector $\boldsymbol{L}$ (of the pump pulse) into the sample plane are parallel. **c**, **d**, Obstructing geometries (③ & ④). In these cases, $\boldsymbol{M}$ and $\boldsymbol{L}$ are antiparallel.

## Photonic OAM-driven magnetization dynamics

Fig. 2 compares the (smoothed) magnetic responses of the sample measured and extracted with the method described above for irradiation of $|l|$ = 5 (Fig. 2**a**) and $|l|$ = 7 (Fig. 2**b**) with the response for $l$ = 0 illumination (black). The supporting (blue) and obstructing (red) cases clearly show a different behavior from one another as well as from the $l$ = 0 measurement. At a pump-probe delay of $\Delta t$ = 0 fs, the drop of the magnetization begins similarly. However, with an increasing pulse delay, the curves split up with the supporting case demagnetizing more strongly than the obstructing case. The difference reaches its maximum at $\Delta t \approx$ 120 fs. After this, the two curves reunite again and follow the same trace. Compared to the demagnetization curve without OAM, it looks like both the supporting and the obstructing curves have “bumps” in the area of separation that point in opposite directions. Crucially, the size of these "bumps" increases significantly as the angle of incidence of the OAM pump beam with respect to the sample surface normal is increased (see example data in Supplementary Fig. 6). This clearly shows that the strength of the OAM-induced magnetization dynamics depends on the relative orientation (angle) between **L** and **M**. Repeating our experiment for other OAM orders reveals similar characteristic OAM-induced signatures in the ultrafast demagnetization traces (see Supplementary Figs. 3-5), which we analyze and discuss in more detail below.

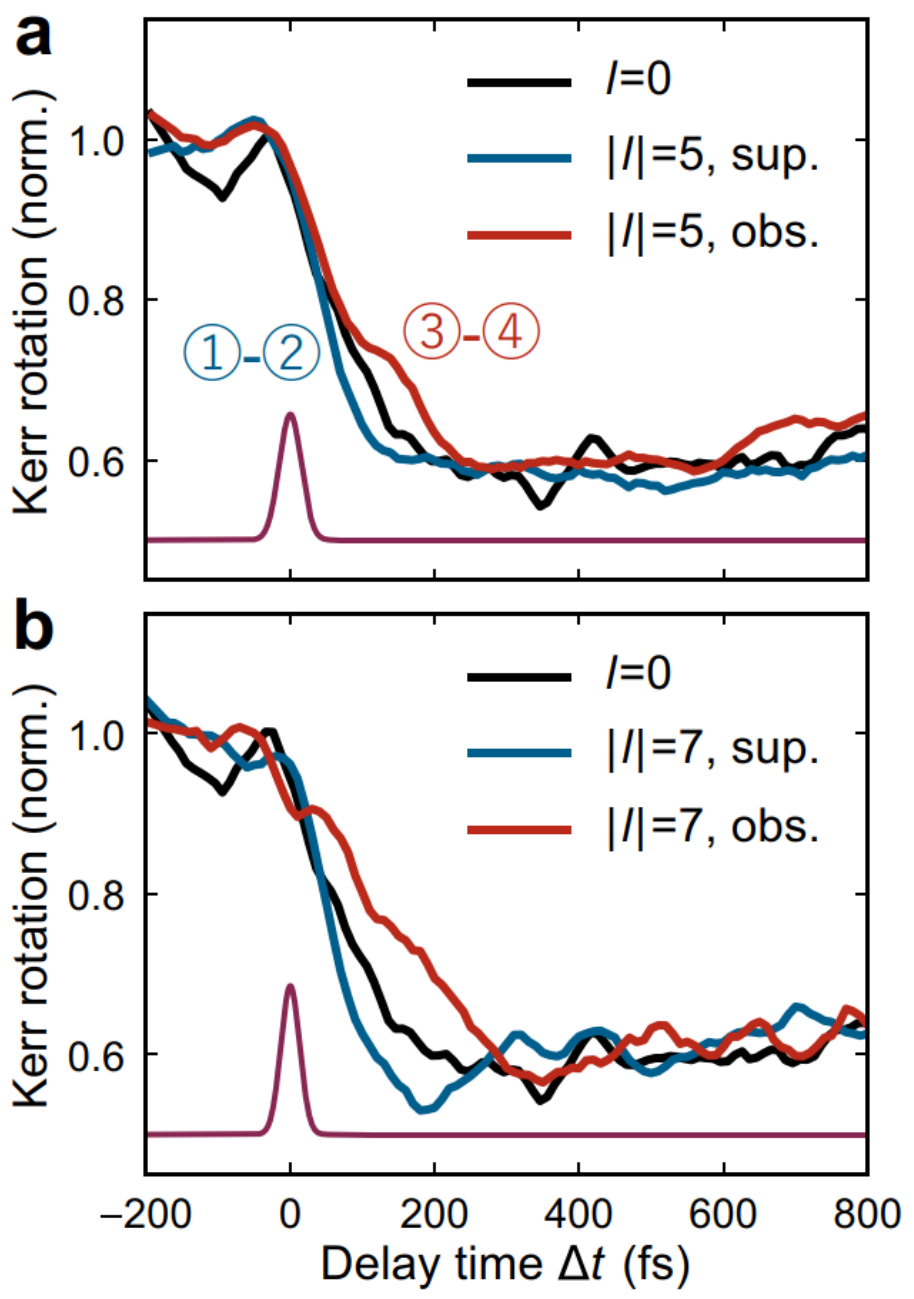

**Fig. 2: Comparison of laser-induced demagnetization with and without photonic OAM.** To best visualize the influence of the OAM, the data points were smoothed and connected with lines. **a**, Demagnetization of the sample after irradiation of a pump pulse with $|l|$ = 5 in the supporting (blue, excitation geometry ① - ②) and obstructing (red, excitation geometry ③ - ④) geometry in comparison to demagnetization with an $l$ = 0 beam (black). **b**, Demagnetization of the sample after irradiation of a pump pulse with $|l|$ = 7 in the supporting and the obstructing geometry in comparison to demagnetization with an $l$ = 0 beam. The purple curves in both plots visualize the respective pump-pulse cross-correlations during the $|l|$ = 5 and $|l|$ = 7 measurements.

Figs. 3**a**, **e** shows the normalized data recorded with the four possible excitation configurations ①, ②, ③ & ④. Figs. 3**b**, **f** shows the sum of the data from the supporting and obstructing geometries (① + ② & ③ + ④), respectively, which contains the nonmagnetic contributions.

These curves consist of a single peak that reaches its maximum approx. at $\Delta t$ = 100 fs. We attribute this to the previously described birefringence of the air molecules in front of the sample that rotate upon strong optical excitation (see also Supplementary Fig. 9). In contrast to the EOKE or SIFE/SOKE signals, this birefringence signal is expected to show a delay with respect to $\Delta t$ = 0 fs since the alignment of the molecules along the pump beam direction is not instantaneous[33,34]. The data in Figs. 3**c**, **g** correspond to the curves already shown in Fig. 2 and are the difference of the supporting and obstructing cases (① - ② & ③ - ④), respectively. Finally, Figs. 3**d**, **h** show the differences of the curves with either an external magnetic field $M^+$ in purple (① - ③) or $M^-$ in green (④ - ②), respectively. These curves visualize the effect that the OAM has for both magnetization directions and hence only contain OAM-dependent contributions. Due to the sequence of the terms in the subtraction, the OAM peaks point in the same direction, in contrast to Figs. 3**c**, **g**. Here, the insets are beam camera images of the pump-probe overlap at the sample location (see methods).

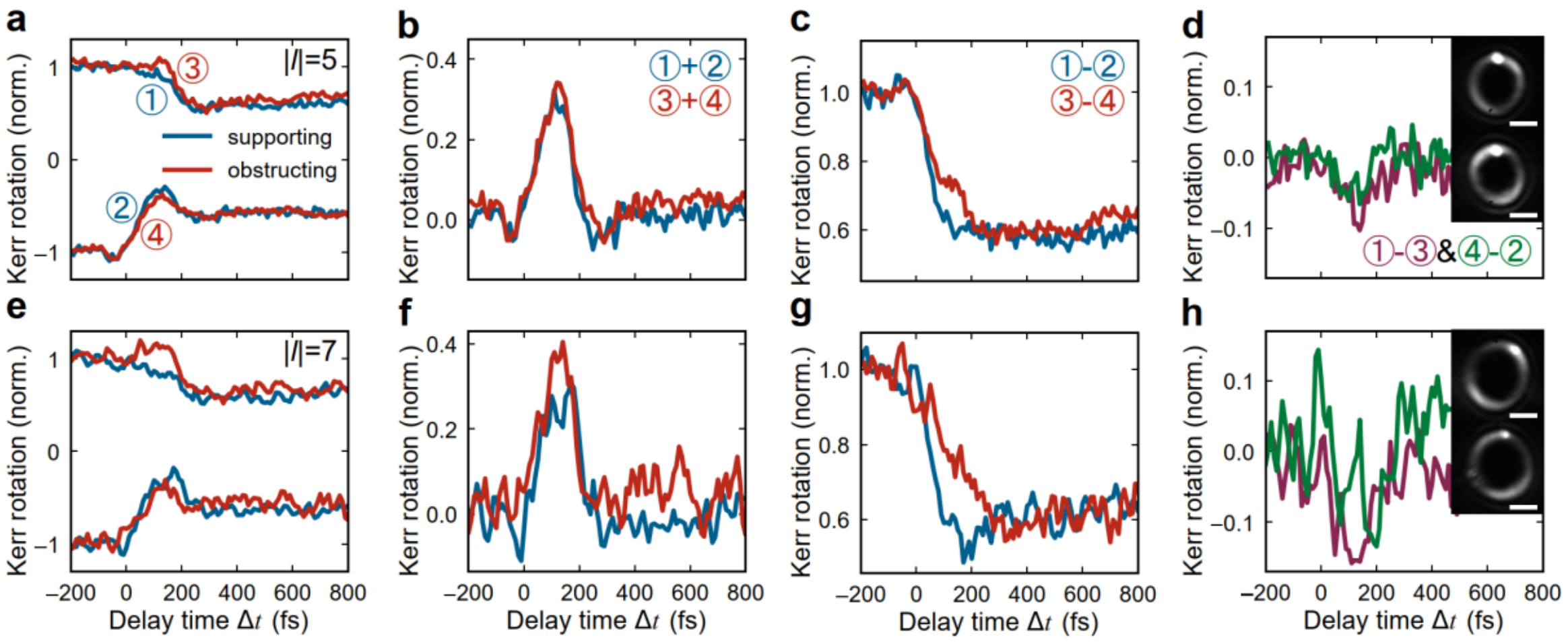

**Fig. 3: Experimental data of |*l*| = 5 (a-d) and |*l*| = 7 (e-h).** Here, the data is not smoothed. For optimal visualization, the data points were connected with lines. The blue lines correspond to data from supporting experimental geometries, while the red curves correspond to data from obstructing geometries. **a**, **e**, Normalized and shifted (in time) experimental data from the four possible excitation geometries. **b**, **f**, Nonmagnetic sum signals (① + ②, blue & ③ + ④, red), calculated from the supporting and obstructing data in **a** and **e**, respectively. **c**, **g**, Magnetic difference signals (① - ②, blue & ③ - ④, red), calculated from the supporting and obstructing data in **a** and **e**, respectively. **d**, **h**, Differences of the data with the same external magnetic fields ($M^+$, ① - ③ & $M^-$, ④ - ②) in purple and green, respectively. The insets are beam camera images that show the pump-probe overlap for $l$ = 5 and $l$ = 7 (top) and $l$ = -5 and $l$ = -7 (bottom), respectively. The scale bars are 1 mm.

Notably, at first glance, it appears like the peaks in the nonmagnetic signal in Figs. 3**b**, **f** occur at similar delay times as the OAM contributions in Figs. 3**c**, **g** and Figs. 3**d**, **h**. In the following, we look into this and conclude that the two phenomena are not connected.

## Fitting model

To quantify the influence of photonic OAM on laser-induced ultrafast demagnetization of nickel, we first focused on the purely magnetic signals (Fig. 2 and Figs. 3**c**, **g**). We developed a simple – effective – fitting model (see methods for a detailed description). In particular, the model's fit functions are not based on a microscopic model and hence only allow us to determine whether the photonic OAM affects the magnetization instantaneously, as would be the case for a primary effect, or with a certain time delay, indicating the involvement of a secondary process.

Our model assumes that the supporting and obstructing cases share a common temporal evolution of the thermally-induced demagnetization similar to the demagnetization with $l = 0$ light and approximates this with a double exponential function (for the de- and remagnetization processes). To determine the influence of the photonic OAM, we model its signature with an additional Gaussian curve that is added to the double exponential function. We assume that this effect is, except for its sign, identical for the supporting and the obstructing cases. Finally, the total magnetic response is convolved with another Gaussian that accounts for the temporal broadening due to the finite durations of the pump and probe pulse. Fig. 4 displays examples of experimental data with different OAM orders that were fitted with this model. The results clearly show that the photonic OAM effect (the dark blue and dark red curves in Fig. 4) is delayed with respect to the beginning of the demagnetization, i.e., it evolves on timescales larger than the temporal width (FWHM) of the pump pulses.

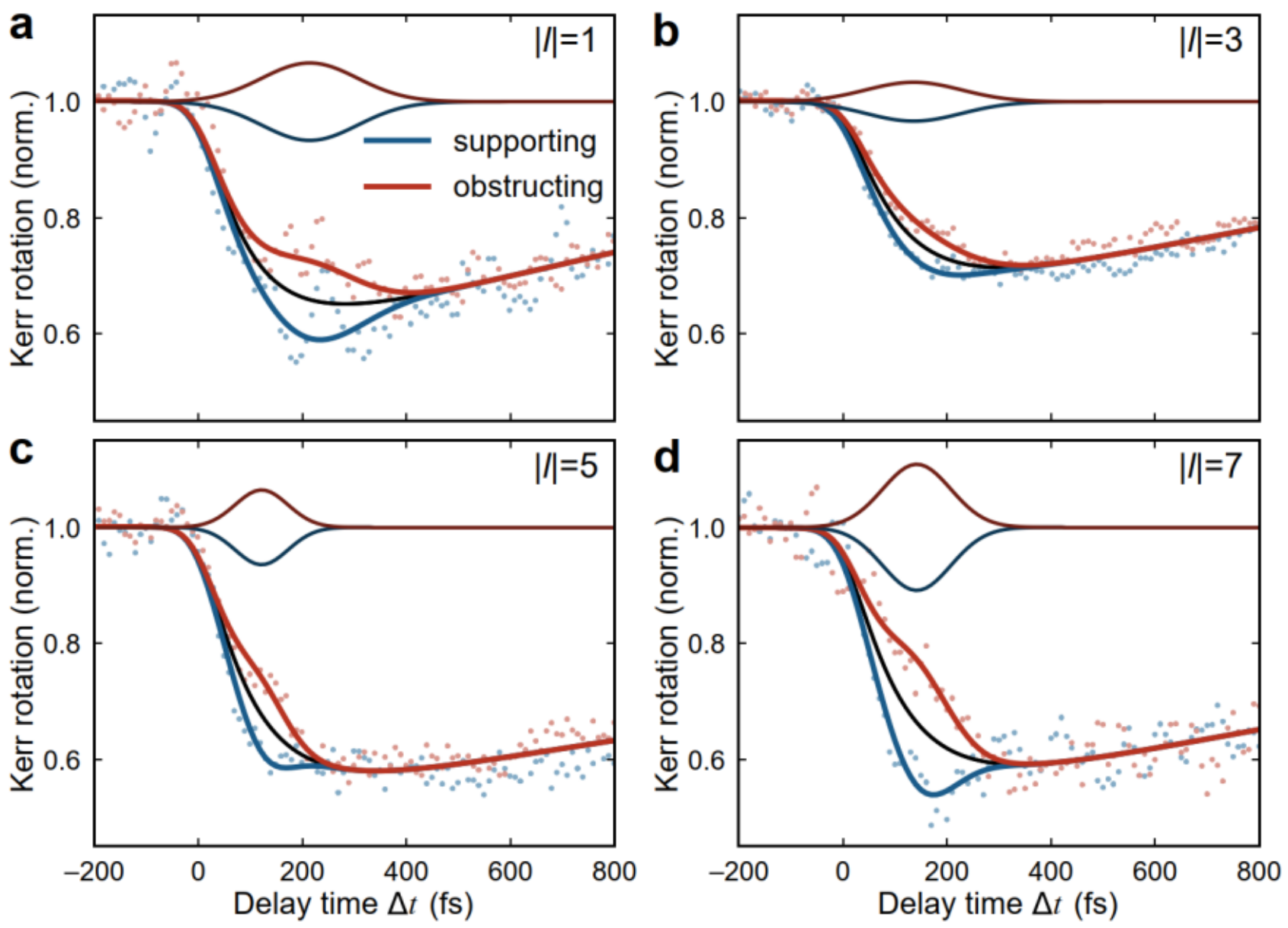

**Fig. 4: Fitting Model.** Fitting of laser-induced demagnetization for different orders of photonic OAM. **a**, $|l| = 1$, **b**, $|l| = 3$, **c**, $|l| = 5$, and **d**, $|l| = 7$. The light blue and light red dots represent the experimental data for the supporting and the obstructing case, respectively. The black lines correspond to the common demagnetization behavior, while the dark blue and red lines represent the additional Gaussians that are added to the shared behavior to account for the supporting/obstructing influence of the photonic OAM (with an offset of +1 for better visualization). The bright blue and red lines show the combinations of the shared behavior and the OAM influence.

**The role of the OAM order for the fs-magnetization dynamics**

Our model allows us to determine how the magnetic response of the sample depends on the order of the photonic OAM of the incident light pulses. For this, the most interesting parameter is the center position of the Gaussians $t_{center}$, which accounts for the OAM influence on the demagnetization (the dark blue and dark red lines in Fig. **4**). This is the time at which the OAM-dependent signal reaches its maximum. In Fig. **5a**), we plotted the relative position of $t_{center}$ with respect to the momentum of the optical excitation $t_0$ from all our measured datasets with different OAM orders as purple circles (see Supplementary Fig. 7 for all datasets not shown in Fig. **4**) against the order of the incident OAM $|l|$.

Our results strongly suggest that the OAM contribution to the demagnetization process reaches its maximum more quickly as the OAM order of the incident light increases, with a difference of more than 100 fs between |l| = 1 and |l| = 8. These findings are confirmed by an independent peak-fitting analysis of data sets containing only OAM-dependent signal contributions (i.e., Figs. **3d, h**, and Supplementary Fig. 10 for the results). The average center positions for both magnetization directions (green squares in Fig. **5a**) are nearly identical to those from our analysis of the magnetic data. Crucially, all these center times are considerably larger than the pump pulse duration ($\tau$pump $\leq$ 35 fs). This clearly shows that the influence of the photonic OAM on the demagnetization dynamics must be caused by a secondary process that depends on the order of the photonic OAM.

In contrast, the nonmagnetic contribution to the Kerr signal in our experiment is independent of the OAM order l. This is illustrated in Fig**. 5a**), which shows the centers of the nonmagnetic peak in the Kerr signal as gray triangles. These times were obtained by fitting the sum signals (e.g., Figs. 3b, f) with single Gaussians (see Supplementary Fig. 8 for the individual fits and Supplementary Fig. 12 for justification of our fitting model). The center times are consistently around $\Delta t$ = 100 fs, supporting our claim that they are solely caused by pump-induced birefringence of the air molecules in front of the sample surface, as discussed above.

Furthermore, we find a clear, material-specific dependence of photonic OAM on the magnetic response by studying the OAM-dependent ultrafast magnetization dynamics of a thin film of another ferromagnetic material, namely the ferromagnetic alloy permalloy (Ni80Fe20, amorphous alloy). Fig. **5b** shows the laser-induced demagnetization of permalloy for |l|=±6 in supporting and obstructing geometries. In contrast to Ni, we find identical magnetization traces for both excitation geometries, indicating a negligible impact of photonic OAM on the ultrafast magnetization dynamics of permalloy. This striking difference between Ni and permalloy provides the foundation for the discussion of the microscopic origin of the indirect conversion of photonic OAM to the spin system.

To support the conclusion of a material-specific, indirect OAM conversion from photons to the spin system, we model the ultrafast magnetization traces with a toy model in which the photonic OAM can be stored in a yet-unknown subsystem L(t) before being transferred to the spin system. The transfer processes themselves are described by simple rate equations with characteristic exchange rates. The model is schematically illustrated in Fig. **5c**. For simplicity, we consider only two subsystems, namely the magnetic subsystem m(t) and a second subsystem L(t) that is capable of storing the photonic OAM inside the material. The ultrafast demagnetization due to laser heating (i.e., the intensity $|E|^2$ of the laser pulse) is fully characterized by the demagnetization and remagnetization times $\tau_D$ and $\tau_R$ and is not directly affected by the OAM of the laser pulse. The photonic OAM $L^{\pm}$ is only deposited in the L subsystem and can unidirectionally be transferred into the magnetic system m(t) by the exchange rate $\frac{1}{\tau_T}$. Crucially, we also consider a so-called accumulation time $t_A$ for the population in the L subsystem, which accounts for the delayed build-up of OAM inside the solid, e.g. due to an indirect transfer of photonic OAM from the light field into the solid. More details on our model can be found in the Supplementary Methods section.

A possible solution of our rate equation model is plotted as solid lines onto the experimental Kerr traces (data points) for the supporting (blue) and obstructing (red) case with $|l| = 5$ in Fig. **5d**. In addition, the corresponding population of the OAM bath L(t) and the Gaussian signature of the OAM contribution to the magnetic signal are shown as green and light red solid lines, respectively. In general, our model is only able to describe the anomalous lineshape of the magnetization traces with photonic OAM when a non-vanishing accumulation time for the OAM population in the material is considered. For the case of $|l| = 5$, the best fit was obtained for an accumulation time of about 70±20 fs and a finite exchange rate $\frac{1}{\tau_T} = (25 \pm 25\ \text{fs})^{-1}$(see Supplementary Figs. 12 and 13, and Supplementary Methods in the Supplementary Information for an in-depth discussion of all model parameters). This indicates that the photonic OAM cannot directly affect the magnetic order inside Ni, but first needs to be accumulated for a few tens of fs inside the solid before gradually affecting the magnetization dynamics of the ferromagnet. Yet, the microscopic origin and interpretation of the OAM storage in the bath L(t) remain elusive.

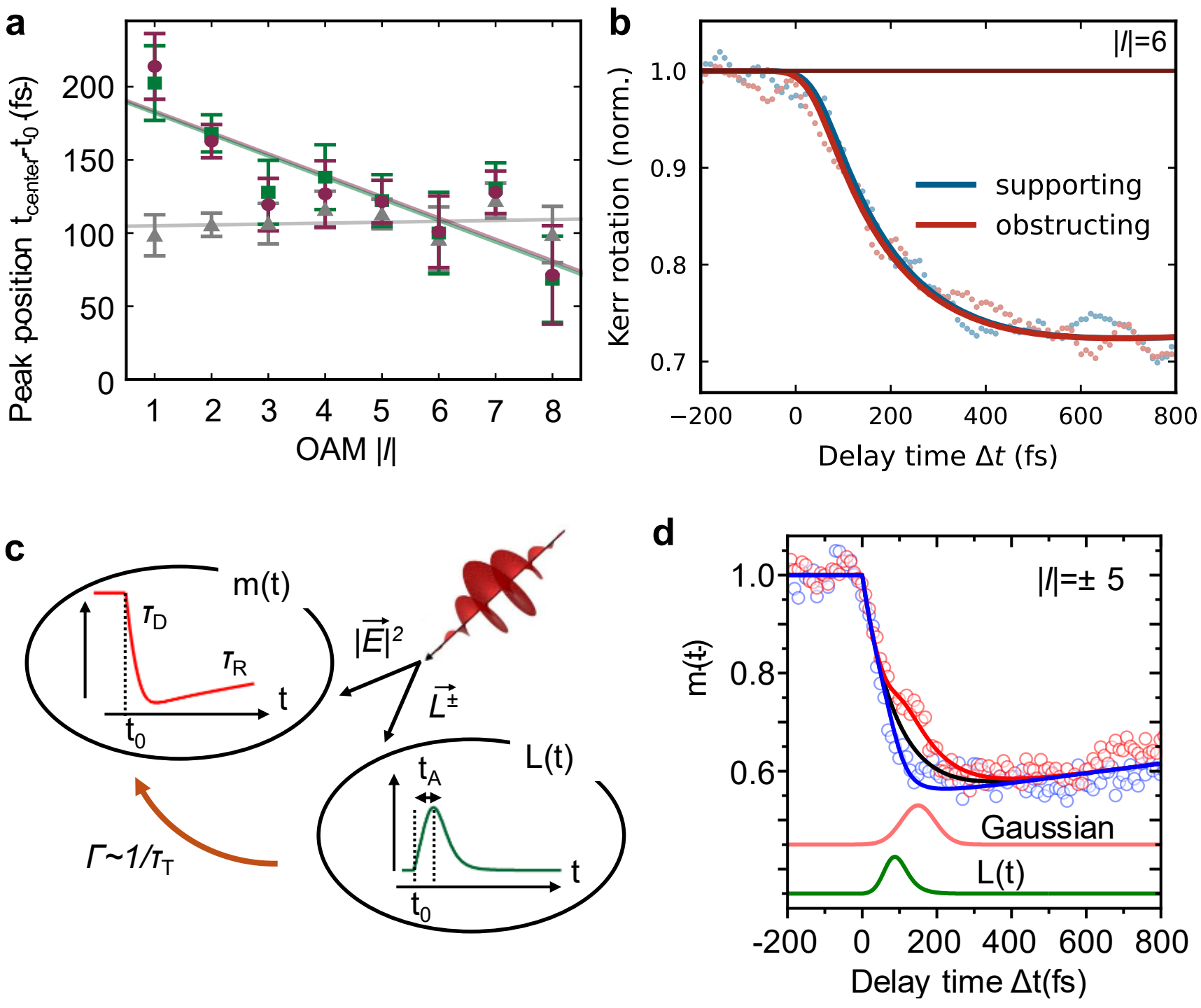

**Fig. 5: Correlation between the photonic OAM and its influence on the demagnetization process. (a)** The purple data points (circles) in this plot were obtained from the fitting model shown in Fig. 4, which is based on equations (4) & (5). They correspond to the time between the values for $t_0$ and the centers of the Gaussians that simulate the OAM influence on the demagnetization for different OAM orders. The green data points (squares) are average values of the centers of Gaussians used to fit data that only contains OAM contributions (i.e., Figs. 3**d**, **h**, equation S9). The grey data points (triangles) are the average center positions of the fitted nonmagnetic peaks [i.e., Figs. 3**b**, **f**, equation (9)]. The lines are linear fits to the data points, intended to serve as guides to the eye. The error bars are based on the fitting errors. Note that $t_0$ is a fit parameter in our model. Its value usually deviates by less than 1 fs from $\Delta t = 0$ fs. Therefore, we plotted $t_{center} - t_0$ on the y-axis. **(b)** Measurement of Py with an OAM of $|l| = 6$. **(c)** Schematic representation of rate-equation model for describing the anomalous lineshape of the magnetization traces m(t) with photonic OAM. The ultrafast demagnetization due to laser

heating (i.e., the intensity $|E|^2$ of the laser pulse) is described by a demagnetization and remagnetization times $\tau_D$ and $\tau_R$. The photonic OAM $L^{\pm}$ is deposited in the L subsystem and its evolution inside the material is described by the called accumulation time $t_A$ and the unidirectional exchange rate $\frac{1}{\tau_T}$ that transfer the photonic OAM from the OAM bath into the magnetic system m(t). More details on our model can be found in the Supplementary Methods section. **(d)** Comparison of the experimental magnetization dynamics for the supporting (light blue dots) and obstructing (light red dots) cases ($|l| = 5$) and our rate equation model (solid blue and red lines). We had to consider a storage time of about 70 ± 20 fs for the OAM population in the material to model the temporal evolution of the experimental magnetization traces. The other model parameters used for computing the model curves are: $\tau_D$= 90 fs, $\tau_T$= 25 fs, $t_A$=70 fs, $\sigma$=25 fs, $L_0$=0.12, $M_0$=0.47, $\tau_R$= 4000 fs.

## Discussion

Our findings for the photonic OAM-driven ultrafast demagnetization dynamics are different from earlier results for the spin of light (i.e., for circularly polarized light). We find conclusive evidence that the photonic OAM manipulates the magnetization dynamics of selected ferromagnets by a secondary process. In contrast, Koopmans et al.[18] demonstrated theoretically, and Dalla-Longa et al.[19] observed experimentally that there is neither a measurable primary[18,19] nor secondary[19] effect of light with spin on the magnetization dynamics. Specifically, regarding a primary process, these studies showed that the photon absorption probability per nickel atom at a realistic pump fluence is too small for a relevant direct transfer of spin from the photons to the electrons, contributing at most ± 0.01 % to the demagnetization. Therefore, according to the calculations from Ref.[18], an OAM of more than $100\hbar$ per photon would be necessary for a measurable influence of the OAM of light in a primary process (assuming that it was possible to transfer this amount of angular momentum in a single absorption process). This fits well with our experimental results that do not show an instantaneous effect, as it would be expected if the electronic transitions were influenced by a primary process like a direct transfer of angular momentum (like an OAM-induced SIFE/SOKE, or, in general, the inverse Faraday effect[36-38]) or dichroism. In the latter case, we would also expect a stronger or weaker overall demagnetization for the supporting and obstructing geometries, respectively, instead of a delayed effect and decays after less than 500 fs. Instead, our observations suggest the involvement of secondary processes that store and subsequently transfer the photonic OAM into the magnetic system of Ni, and that are suppressed, or at least substantially weakened, in permalloy.

There are several mechanisms that could be responsible for the storage and indirection conversion of photonic OAM to the spin system. A first candidate is the existence and gradual buildup of circular currents that are generated by the OAM of light. When induced by circularly polarized light, this phenomenon is called the circular photogalvanic effect[39,40]. With photonic OAM, it has been observed experimentally for the type-II Weyl semimetal $WTe_2$[41] and theoretically predicted for semiconductors[42] as well as plasmas[43]. This phenomenon has also been discussed as a potential reason for the OAM light-induced change in the magnetic anisotropy of an interlayer exchange coupling system[27] with the current inducing a magnetic field. However, so far, such a photocurrent has, to our knowledge, neither been predicted nor observed for simple, elemental, and centrosymmetric ferromagnetic systems such as nickel without spin-momentum locking of the electrons and is thus a rather unlikely explanation.

A second and far more relevant effect is the generation of mesoscopic orbital polarizations. In this scenario, photonic OAM is first converted into orbital polarization and thus stored in the orbital degree of freedom of itinerant electrons in Ni. Such orbital angular momentum of itinerant electrons can be generated even for small SCO, as demonstrated by the orbital Hall effect in Ti[44]. In addition, laser excitation of Ni can lead to orbital currents[45,46] in Ni, which underlines the crucial role of the orbital degree of freedom in this material. Along these lines, we propose that photonic OAM is stored in the orbital degree of freedom and the orbital polarization (corresponding to the bath L(t)), and is subsequently converted into the spin degree of freedom. This mechanism is also fully consistent with the substantially different photonic OAM conversion in Ni and permalloy. On the one hand, Ni has been predicted to show the strongest orbital-to-spin conversion[47], and thus orbital currents can be very efficiently converted into spin currents in Ni[48]. In contrast, orbital-torque experiments on Ti-based heterostructures found only vanishingly small conversion for a similar Ni-Fe alloy[49].

In conclusion, we have demonstrated that the OAM of light significantly influences the ultrafast demagnetization dynamics of a thin nickel film but not of a permalloy film. The conversion of photonic OAM to the spin degree of freedom is delayed relative to the excitation of the system and can, therefore, be attributed to a secondary process. Depending on the measurement geometry, the OAM temporarily either supports or obstructs the demagnetization process for a time span of hundreds of femtoseconds. We also showed that the effect occurs more rapidly with increasing OAM order. Based on the material-specific magnetic response of Ni and permalloy to photonic OAM, we propose that photonic OAM is stored in the orbital degree of freedom and subsequently converted into the spin system.

Thus, our results have introduced a new paradigm for manipulating and controlling magnetization and spin order by OAM light. However, to fully exploit the potential of photonic OAM in the field of fs magnetization, a comprehensive and microscopic understanding of the OAM conversion from the fs light pulses to the spin system is essential. This requires future theoretical and experimental efforts to predict microscopic models for the observed secondary process and to confirm and challenge these predictions by experiments with, for example, different photon energies, angles of incidence, or sample systems.

# Methods

## Experimental setup

In our experiments, to create ultrashort ($\leq$ 35 fs) laser, we used a Ti:sapphire amplifier (Dragon from KMlabs) with a central wavelength of 800 nm and a repetition rate of 6 kHz and a Ti:sapphire amplifier (Astrella from Coherent) with a central wavelength of 800 nm and a repetition rate of 1 kHz. For the pump-probe measurements, each pulse was split into two pulses, which were delayed with respect to each other in 10 fs steps with a Mach-Zehnder interferometer. We frequency-doubled the probe beam with a $\beta$-barium borate (BBO) crystal to a central wavelength of 400 nm before focusing it onto the sample with a pulse length of $\leq$ 75 fs. We sent the pump beam through a commercially available spiral phase plate (VPP-m780 from RPC photonics) with the orders 1 – 8, adding OAM to the pulses. We experimentally confirmed the created OAM orders by focusing the beam with a cylindrical lens and recording beam camera images close to the focus, see Supplementary Fig. 1. The intensity distributions at this position exhibit a number of tilted dark stripes. The number of these stripes equals the total amount of OAM, while the tilt direction depends on the OAM handedness[50]. Both beams were linearly polarized and their intensities were adjusted with combinations of half-wave plates and linear polarizers.

We performed the TR-MOKE experiments with the setup that is schematically depicted in Supplementary Fig. 2. We used a longitudinal geometry with an angle of incidence of ~ 45° of both beams with respect to the sample normal. The angle difference between the two beams is less than 1.65°. Considering the diameter of the probe spot size of approx. 500 µm, this leads to a time smear of less than 48 fs (absolute width, not FWHM) which is below the cross-correlation FWHM of the pump and probe pulses of $\sigma$= 83 fs. To monitor changes in the sample magnetization, we measured changes in the polarization of the probe beam after reflection from the sample. An external magnetic field was applied to saturate the sample. A flip mirror allowed us to monitor the overlap of the pump and probe beams with a beam camera in sample distance before each measurement (see the insets in Fig. 3 and Supplementary Figs. 3, 4, and 5).

## Data evaluation

For each data set with photonic OAM, we performed four time-resolved measurements of the different combinations of sample magnetization direction and incident OAM handedness, respectively. While the sample magnetization could easily be changed by switching the direction of the external magnetic field, changing the OAM handedness required turning the spiral phase plate by 180°. Since the latter made it necessary to readjust the beam overlap, this could lead to a change in time zero. To account for this, we compared the sum signals of the two measurements with the same OAM handedness, respectively, which correspond to nonmagnetic signal contributions, and matched them by shifting the data corresponding to one OAM handedness in time. The first columns in Supplementary Figs. 3-5 show the (normalized) resulting data (note that Supplementary Fig. 5i-l shows data with $l$ = 0, which does not have supporting and obstructing excitation geometries).

The measurements with different OAM handedness can only be used for further evaluation if they were performed with similar fluences in the probed regions. Before the measurements, we optimized this with the beam camera (see insets in the fourth columns of Fig. 3 and Supplementary Figs. 3-5; the brightness of the images can vary due to a grey filter wheel that had to be manually adjusted before each measurement). During the evaluation, we verified the quality of our adjustment by calculating and comparing the sum signal of the supporting and obstructing excitation geometries, respectively. These nonmagnetic signal contributions (second columns of Fig. 3 and Supplementary Figs. 3-5) should ideally be identical, especially after the initial peak.

To see if the photonic OAM influences the demagnetization behavior, we calculated the difference signal of the supporting and obstructing geometries, respectively (see the third columns in Fig. 3 and Supplementary Figs. 3-5).

The fourth columns in Fig. 3 and Supplementary Figs. 3-5 show the differences of data with the same external magnetic field direction but different incident OAM handednesses. The data resulting from this should show all effects of the photonic OAM, both magnetic and nonmagnetic, but without all OAM-independent contributions like the thermal demagnetization and the nonmagnetic refractive index change of the air in front of the sample. For the $l$ = 0 measurements, Supplementary Fig. 51 shows the magnetic difference signals fitted with equation (7).

**Fitting model (demagnetization)**

Our fitting model for the ultrafast demagnetization with OAM light is composed of three different contributions:

1. The "normal" demagnetization without the OAM influence with a single exponential decay followed by an exponential increase:

$$s_{demag}(\Delta t) = M_0 \cdot (-\Theta(\Delta t - t_0) + 1) + \left( M_{relax} + (M_0 - M_{min}) \cdot e^{-\frac{\Delta t - t_0}{\tau_{demag}}} + (M_{min} - M_{relax}) \cdot e^{-\frac{\Delta t - t_0}{\tau_{relax}}} \right) \cdot \Theta(\Delta t - t_0) \quad (1)$$

with $\Delta t$ the time delay between pump and probe pulse, $M_0 = 1$ the initial (saturation) magnetization, $\Theta$ the Heavyside step function, $t_0$ the temporal overlap of pump and probe pulse (time zero), $M_{min}$ the magnetization to which the exponential decay converges, $\tau_{demag}$ the time constant of the exponential decay, $M_{relax}$ the magnetization to which the exponential increase converges, and $\tau_{relax}$ the time constant of the exponential increase.

2. The influence of the photonic OAM on the demagnetization process which we model with a Gaussian:

$$s_{OAM}(\Delta t) = \left( \frac{S}{\sqrt{2\pi}} \cdot \frac{2\sqrt{2\ln(2)}}{\sigma_{OAM}} \cdot e^{-\frac{(\Delta t - t_{center})^2 \left(2\sqrt{2\ln(2)}\right)^2}{2\sigma_{OAM}^2}} \right) \cdot \Theta(\Delta t - t_0) \quad (2)$$

with $S$ a scale factor, $\sigma_{OAM}$ the FWHM, and $t_{center}$ the center position of the Gaussian.

3. The experimental time smear due to the finite lengths of pump and probe pulse:

$$s_{time_smear}(\Delta t) = \frac{1}{\sqrt{2\pi}} \cdot \frac{2\sqrt{2\ln(2)}}{\sigma} \cdot e^{-\frac{\Delta t^2 \left(2\sqrt{2\ln(2)}\right)^2}{2\sigma^2}} \quad (3)$$

with $\sigma = 83$ fs the FWHM. This value was determined experimentally from an XFROG trace of a pump pulse without OAM and a probe pulse. $s_{time_smear}$ is normalized to $s_{time_smear,norm}$ for all further calculations.

For fitting the supporting contributions, we calculate the difference

$$s_{sup}(\Delta t) = s_{demag}(\Delta t) - s_{OAM}(\Delta t), \quad (4)$$

while for the obstructing contributions, we calculate the sum

$$s_{obs}(\Delta t) = s_{demag}(\Delta t) + s_{OAM}(\Delta t)\,. \tag{5}$$

To account for the time smear, we convolve both functions with $s_{time_smear,norm}(\Delta t)$, respectively:

$$s_{sup/obs,time_smear}(\Delta t) = s_{sup/obs}(\Delta t) \circ s_{time_smear,norm}(\Delta t)\,. \tag{6}$$

Finally, we simultaneously fit $s_{sup,time_smear}(\Delta t)$ and $s_{obs,time_smear}(\Delta t)$ with the same fitting parameters. These functions correspond to the blue and red curves in Fig. 4 and Supplementary Fig. 7, respectively. The black curves in these figures are given by $s_{demag}(\Delta t) \circ s_{time_smear,norm}(\Delta t)$ while the dark blue and dark red curves are given by $1 + s_{OAM}(\Delta t) \circ s_{time_smear,norm}(\Delta t)$ and $1 - s_{OAM}(\Delta t) \circ s_{time_smear,norm}(\Delta t)$, respectively. In Supplementary Fig. 11, we plot the minima of $s_{demag}(\Delta t) \circ s_{time_smear,norm}(\Delta t)$ against the corresponding OAM orders for all datasets and find no dependence.

The data with $l = 0$ we fitted with

$$s_{l=0}(\Delta t) = s_{demag}(\Delta t) \circ s_{time_smear,norm}(\Delta t)\,. \tag{7}$$

We show the fitting results in in Supplementary Fis. 5**l**. We provide a complete list of the fitting parameters for all OAM orders in Supplementary Table 1.
We want to stress that $t_0$ is a fit parameter, that usually deviates by less than 1 fs from $\Delta t = 0$ fs, which we achieved by shifting the data in integer fs increments.

**Fitting model (nonmagnetic peaks)**

We fitted the nonmagnetic peaks with simple Gaussians to extract their center positions $t_{center,nonmag}$. The fitting function is:

$$s_{nonmag}(\Delta t) = \left( \frac{S_{nonmag}}{\sqrt{2\pi}} \cdot \frac{2\sqrt{2\ln(2)}}{\sigma_{nonmag}} \cdot e^{-\frac{(\Delta t - t_{center,nonmag})^2 \left(2\sqrt{2\ln(2)}\right)^2}{2\sigma_{nonmag}^2}} \right) \tag{8}$$

with the FWHM $\sigma_{nonmag}$. We show the fitting results in Supplementary Data Fig. 7. We provide a complete list of the fitting parameters in Supplementary Table 2.

**Fitting model (only OAM effect)**

We fitted the peaks of only the OAM effect with simple Gaussians to extract their center positions $t_{center,onlyOAM}$. The fitting function is:

$$s_{onlyOAM}(\Delta t) = \left( \frac{S_{onlyOAM}}{\sqrt{2\pi}} \cdot \frac{2\sqrt{2\ln(2)}}{\sigma_{onlyOAM}} \cdot e^{-\frac{(\Delta t - t_{center,onlyOAM})^2 \left(2\sqrt{2\ln(2)}\right)^2}{2\sigma_{onlyOAM}^2}} \right) \tag{9}$$

with the FWHM $\sigma_{onlyOAM}$. We show the fitting results in Supplementary Fig. 10 and provide a complete list of the fitting parameters in Supplementary Table 3.

**Data availability:** The data supporting the findings of this study are available from the corresponding author upon reasonable request.

**Code availability:** The code for evaluating and fitting the experimental data is available from the corresponding author upon request.

**Acknowledgments:** E. Prinz, P. Herrgen, and J. Hoefer would like to thank Martin Stiehl, Simon Häuser, and Stephan Wust for support with the experimental setup. The experimental work was funded by the Deutsche Forschungsgemeinschaft (DFG, German Research Foundation) - TRR 173 - 268565370 Spin + X: spin in its collective environment (Projects A08 and B03). B. Stadtmüller acknowledges financial support from the Dynamics and Topology Center funded by the State of Rhineland Palatinate.

**Author contributions:** E. P., B. S., and M. A. conceived the experiments and formulated the theoretical model. E. P., P.H., and J. H. performed the experiments. E. P. and P.H. analyzed the data. E.P. wrote the first draft of the manuscript with the help of B.S. and M.A.. All authors worked on the revised manuscript.

**Competing interest declaration:** The authors declare no competing interests.

## Supplementary Information for

# Twisted light affects ultrafast demagnetization

**Authors:** Eva Prinz[1], Paul Herrgen[1*], Jonas Hoefer[1], Benjamin Stadtmüller[1,2], Martin Aeschlimann[1]

**Affiliations:**

[1]Department of Physics and Research Center OPTIMAS, RPTU University Kaiserslautern-Landau, 67663 Kaiserslautern, Germany

[2]Experimentalphysik II, Institute of Physics, Augsburg University, D-86135 Augsburg, Germany.

*Corresponding author. Email: pherrgen@rptu.de

### Supplementary Figures

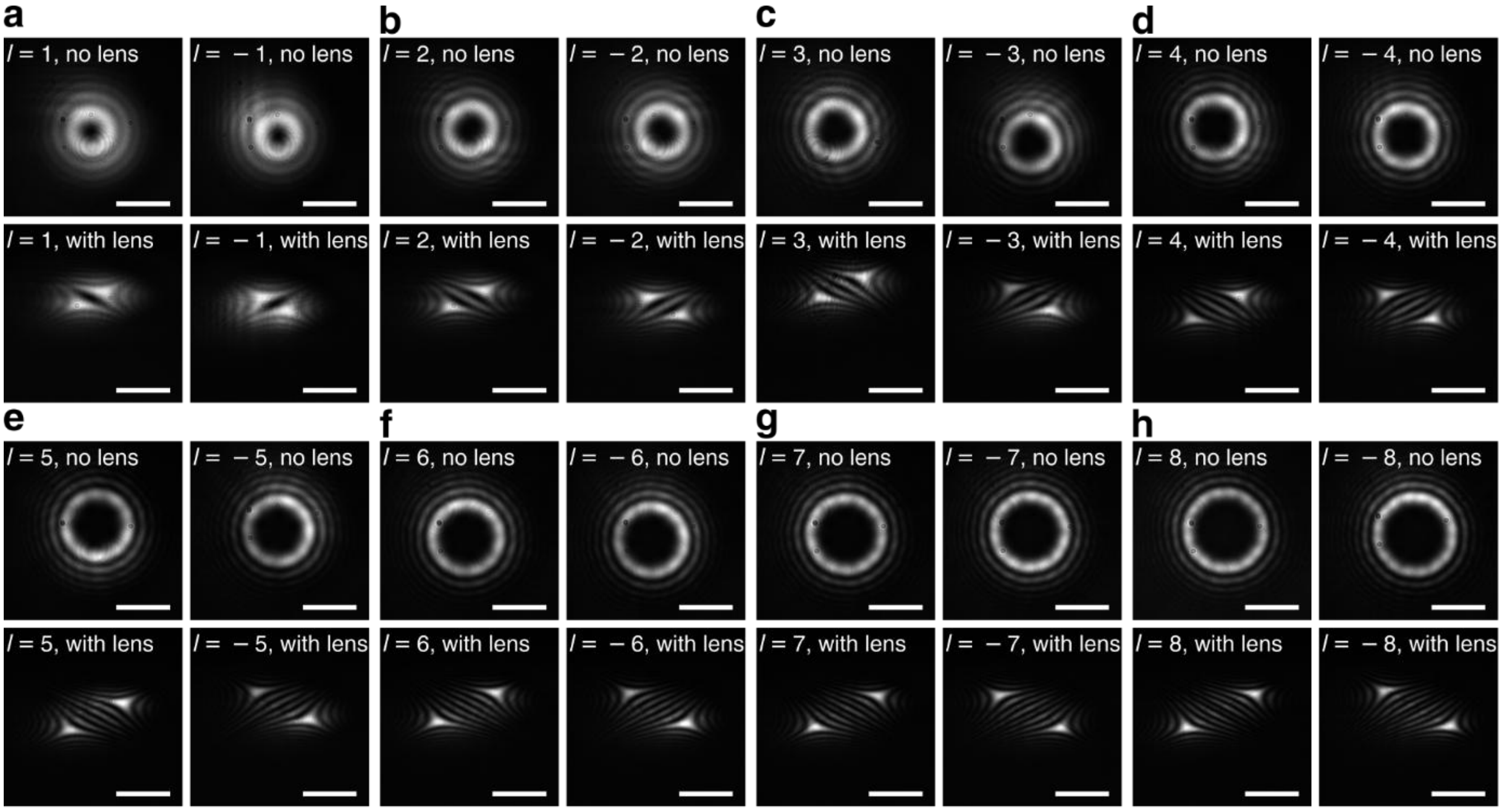

**Supplementary Fig. 1: Beam camera images of different OAM orders.** **a** $|l|=1$, **b** $|l|=2$, **c** $|l|=3$, **d** $|l|=4$, **e** $|l|=5$, **f** $|l|=6$, **g** $|l|=7$, and **h** $|l|=8$. The images in the top rows show the beam profiles of the OAM with a positive sign on the left and a negative sign on the right. The images in the bottom row correspond to the same OAM values but were recorded near the focus of a cylindrical lens.

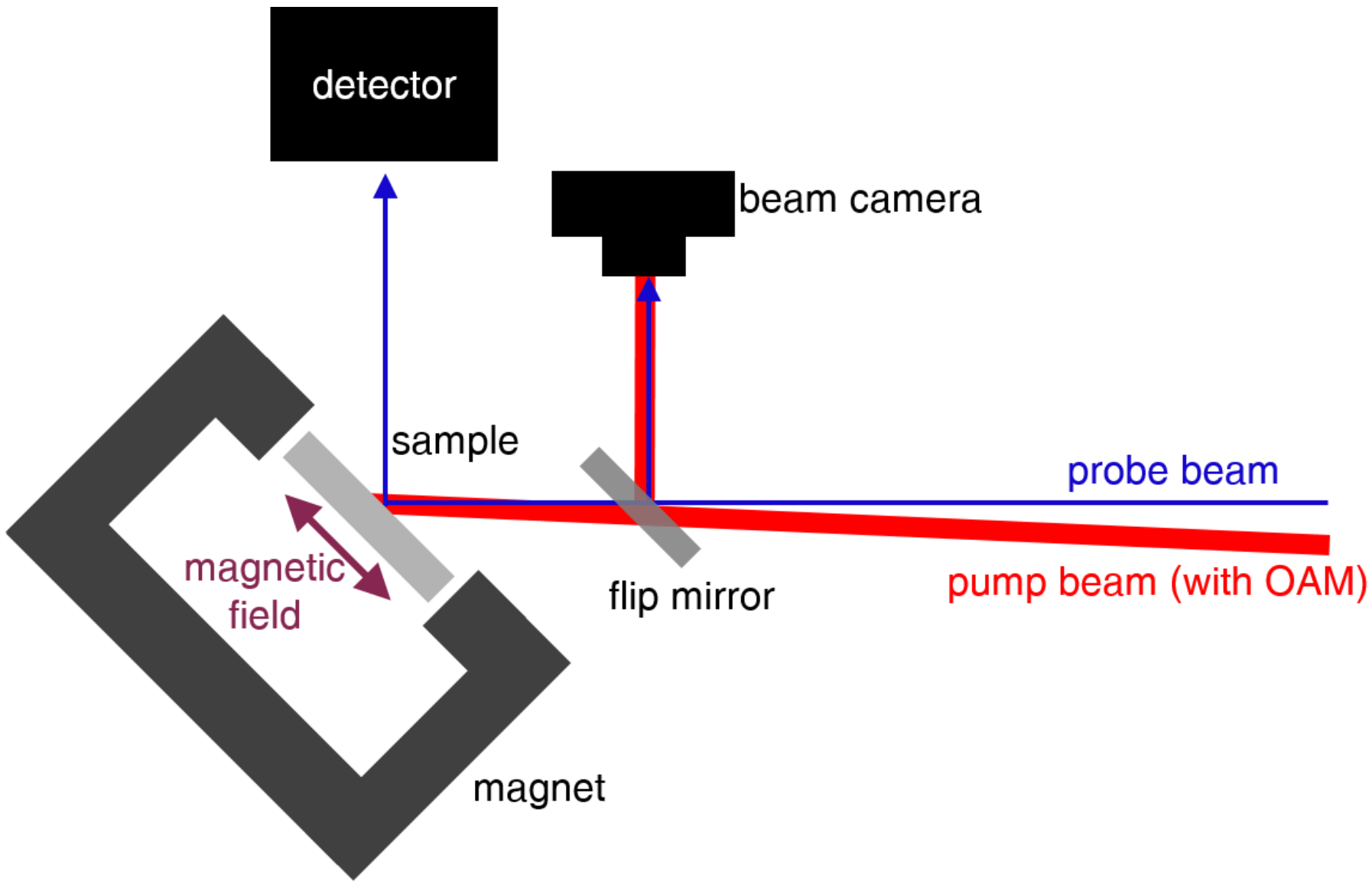

**Supplementary Fig. 2: Schematic of the longitudinal TR-MOKE setup.**

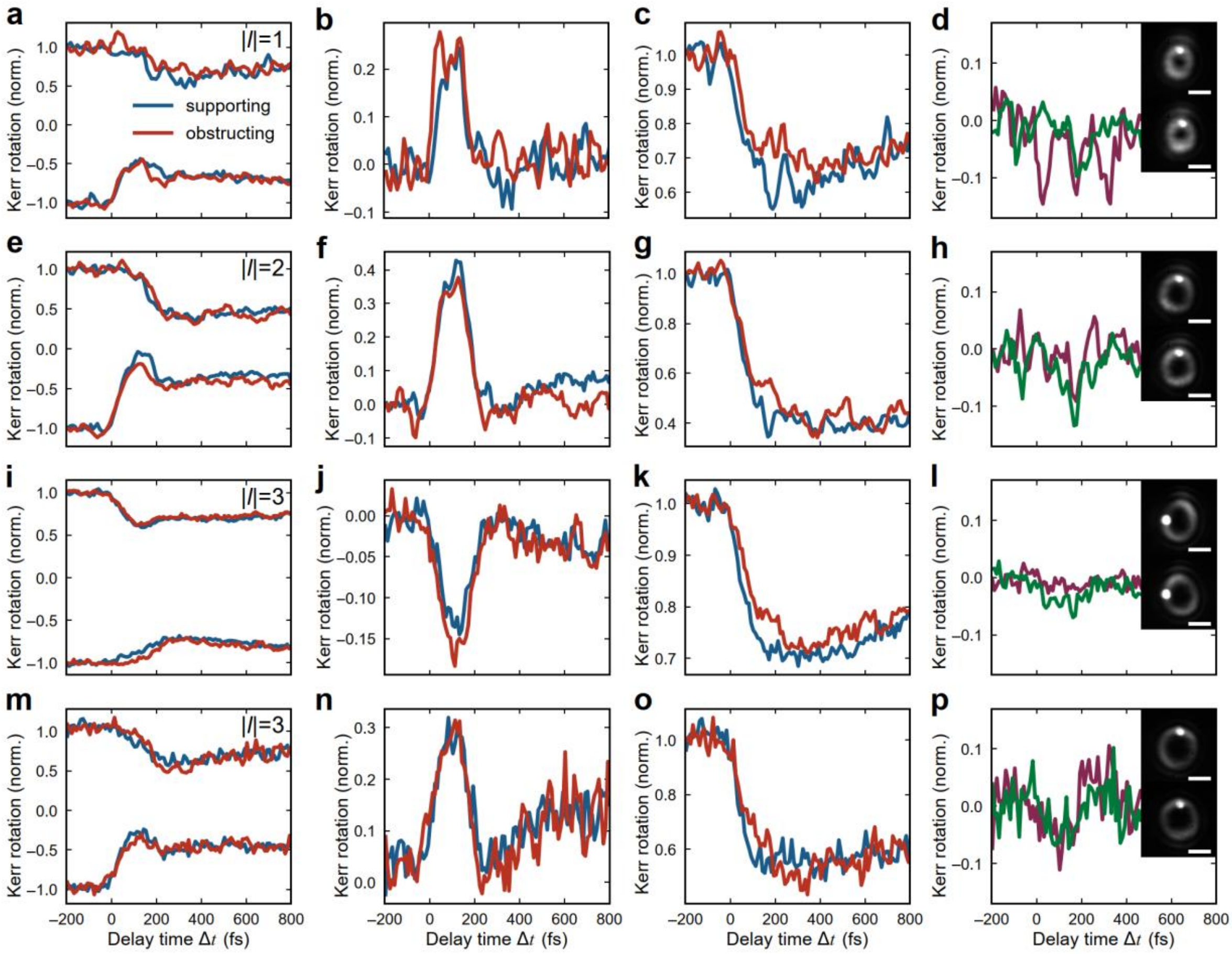

**Supplementary Fig. 3: Experimental data of |*l*| = 1 (a-d), |*l*| = 2 (e-h) and |*l*| = 3 with lower (i-l) and higher pump fluences (m-p).** For optimal visualization, the data points were connected with lines. The blue lines correspond to data from supporting experimental geometries, while the red curves correspond to data from obstructing geometries. **a**, **e**, **i**, **m** Normalized and shifted (in time) experimental data from the four possible excitation geometries. **b**, **f**, **j**, **n** Nonmagnetic sum signal, calculated from the supporting and obstructing data in **a**, **e**, **i**, and **n**, respectively. **c**, **g**, **k**, **o** Magnetic difference signal, calculated from the supporting and obstructing data in **a**, **e**, **i**, and **n**, respectively. **d**, **h**, **l**, **p** Differences of the data with the same external magnetic fields ($M^+$ and $M^-$) in purple and green, respectively. The insets are beam camera images that show the pump-probe overlap for $l$ = 1, $l$ = 2, and $l$ = 3 (top) and $l$ = -1, $l$ = -2, and $l$ =-3 (bottom), respectively. The scale bars are 1 mm.

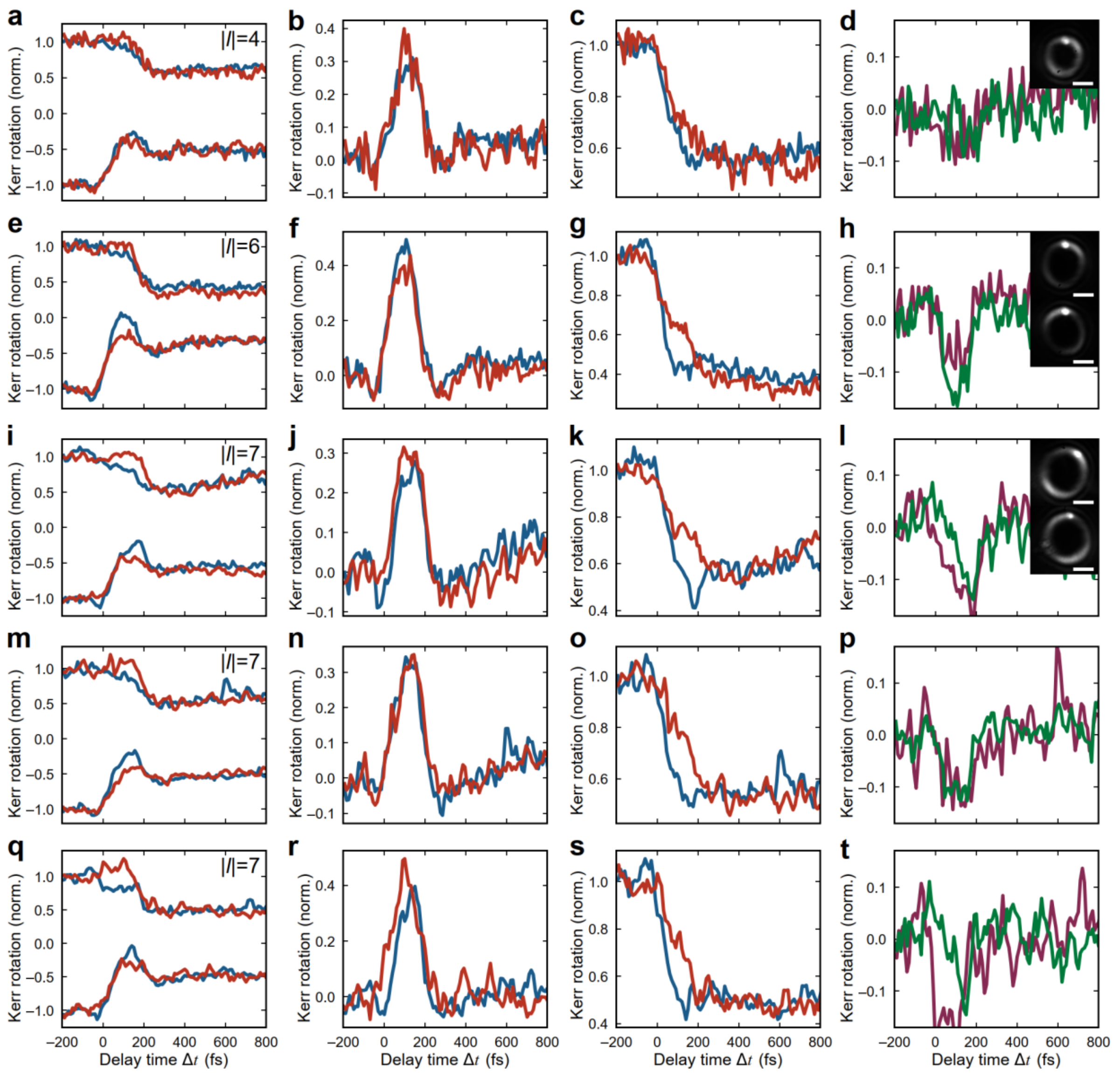

**Supplementary Fig. 4: Experimental data of |*l*| = 4 (a-d), |*l*| = 6 (e-h) and |*l*| = 7 from the lowest pump fluence in (i-l) to the highest pump fluence in (q-t).** For optimal visualization, the data points were connected with lines. The blue lines correspond to data from supporting experimental geometries, while the red curves correspond to data from obstructing geometries. **a**, **e**, **i**, **m**, **q** Normalized and shifted (in time) experimental data from the four possible excitation geometries. **b**, **f**, **j**, **n**, **r** Nonmagnetic sum signal, calculated from the supporting and obstructing data in **a**, **e**, **i**, **m**, and **q**, respectively. **c**, **g**, **k**, **o**, **s** Magnetic difference signal, calculated from the supporting and obstructing data in **a**, **e**, **i**, **m**, and **q**, respectively. **d**, **h**, **l**, **p**, **t** Differences of the data with the same external magnetic fields ($M^+$ and $M^-$) in purple and green, respectively. The inset in **d** is a beam camera image that shows the pump-probe overlap for *l* = 4 (the overlap was also checked for l= -4; however, the image was accidentally not saved). The insets in **h** and **l** are beam camera images that show the pump-probe overlap for *l* = 6 and *l* = 7 (top) and *l* = -6 and *l* = -7 (bottom), respectively. The overlap is the same for all *l* = 7/*l* = -7 measurements, respectively. The scale bars are 1 mm. The data for |*l*| = 5 and the lowest |*l*| = 7 fluence is shown in Fig. 3.

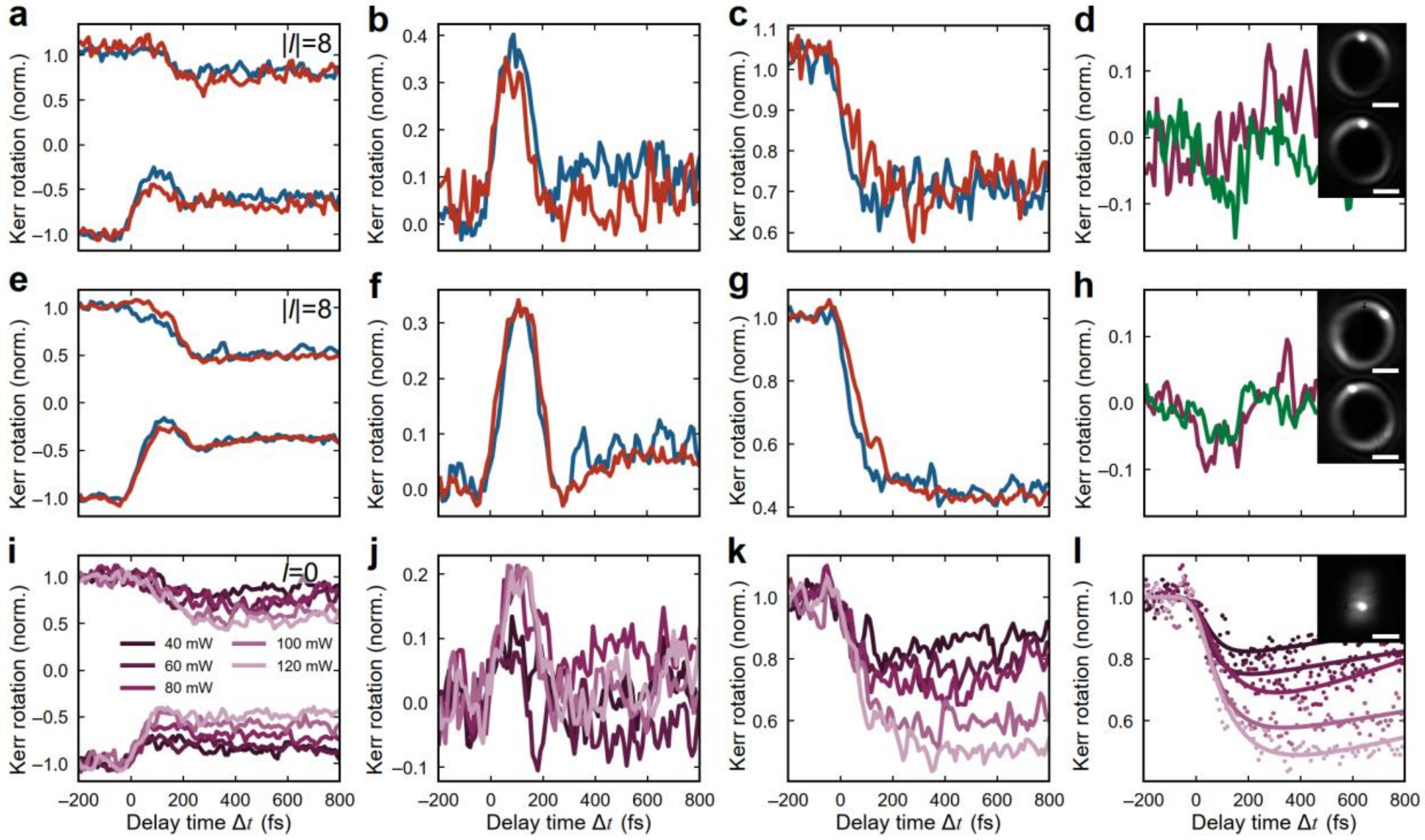

**Supplementary Fig. 5: Experimental data of |*l*| = 8 with lower (a-d) and higher pump fluences (e-h), and *l* = 0 with different pump fluences (i-l).** For optimal visualization, the data points were connected with lines. The blue lines correspond to data from supporting experimental geometries, while the red curves correspond to data from obstructing geometries. **a**, **e**, and **i**, Normalized and shifted (in time) experimental data from the four possible excitation geometries. **b**, **f**, and **j**, Nonmagnetic sum signal, calculated from the in **a**, **e**, and **i**, respectively. **c**, **g**, **k** Magnetic difference signal, calculated from the supporting and obstructing data in **a**, **e**, and **i**, respectively. **d**, **h**, Differences of the data with the same external magnetic fields ($M^+$ and $M^-$) in purple and green, respectively. **l**, Fitting of the data in **k**. The dots represent the experimental data, while the lines correspond to the fits. The insets in **d** and **h** are beam camera images that show the pump-probe overlap for *l* = 8 (top) and *l* = -8 (bottom), respectively. The inset in **l** is a beam camera image that shows the pump-probe overlap for *l* = 0. The scale bars are 1 mm.

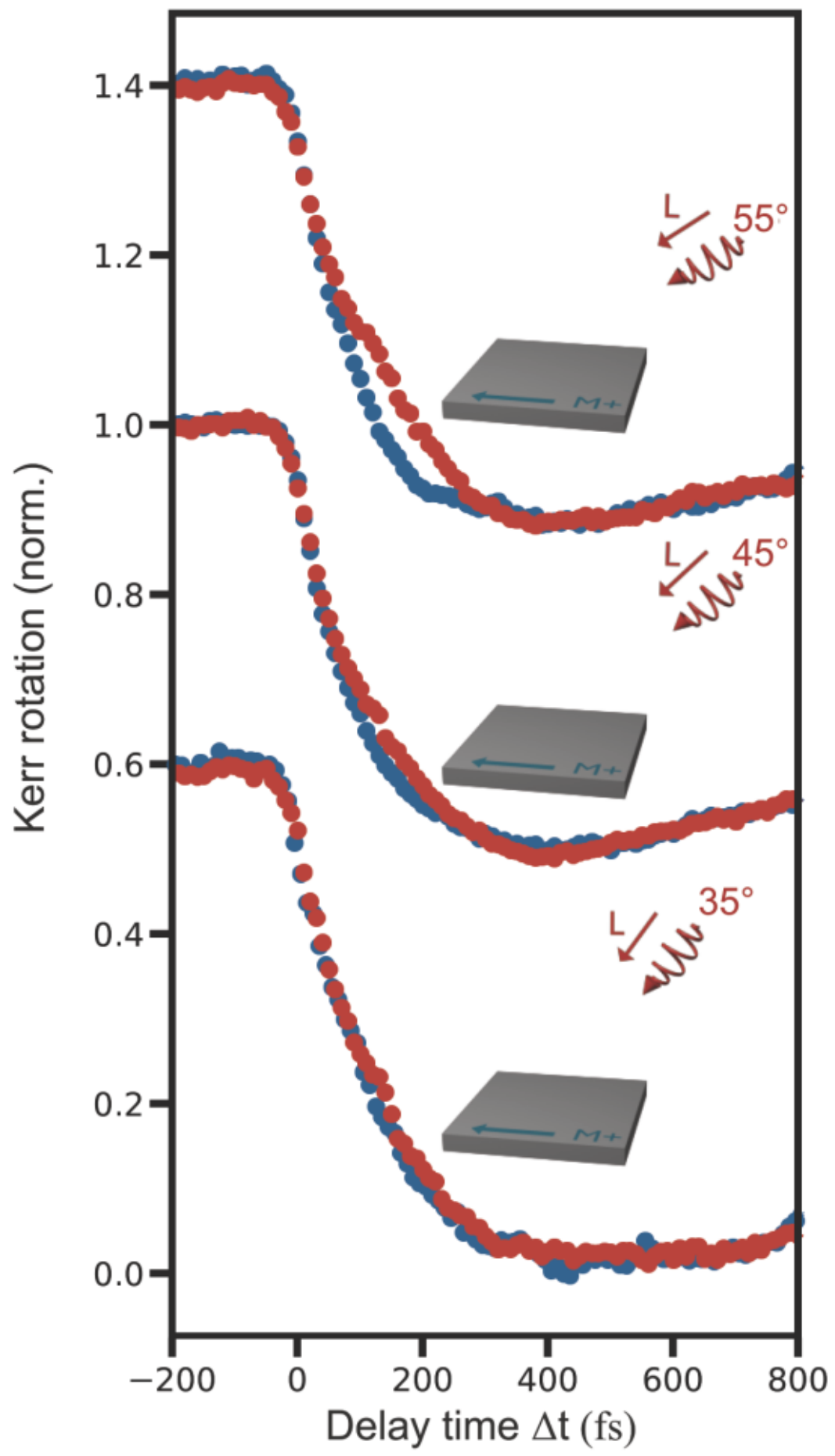

**Supplementary Fig. 6: Experimental data of |*l*| = 7 for different angles of incidence with respect to the sample surfaces.** The blue and light red dots represent the experimental data for the supporting and the obstructing cases, respectively. The insets highlight the change in the experimental geometry for different angles of incidence. We have selected time-resolved Kerr traces with comparable loss of magnetic order for the different experimental geometries.

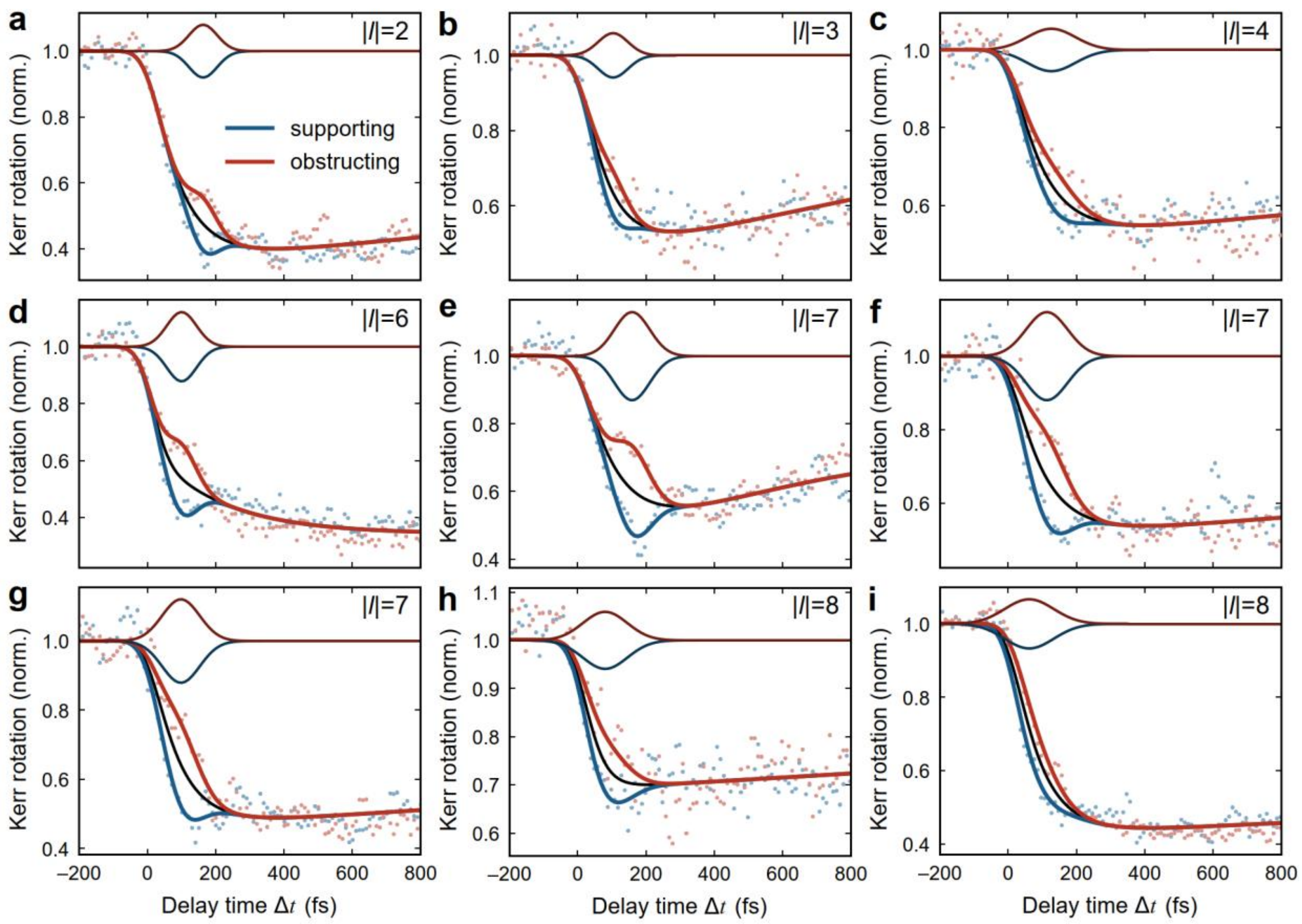

**Supplementary Fig. 7: Fitting of laser-induced demagnetization for different orders of photonic OAM that is not shown in the main manuscript.** **a**, $|l| = 2$, **b**, $|l| = 3$, **c**, $|l| = 4$, **d**, $|l| = 6$, **e-g**, $|l| = 7$ (increasing pump fluences), and **h-i**, $|l| = 8$ (increasing pump fluences). The light blue and light red dots represent the experimental data for the supporting and the obstructing case, respectively. The black lines correspond to the common demagnetization behavior, while the dark blue and red lines are the additional Gaussians that are added to the shared behavior to account for the supporting/obstructing influence of the optical OAM (with an offset of +1 for better visualization). The bright blue and red lines show the combinations of the shared behavior and the OAM influence.

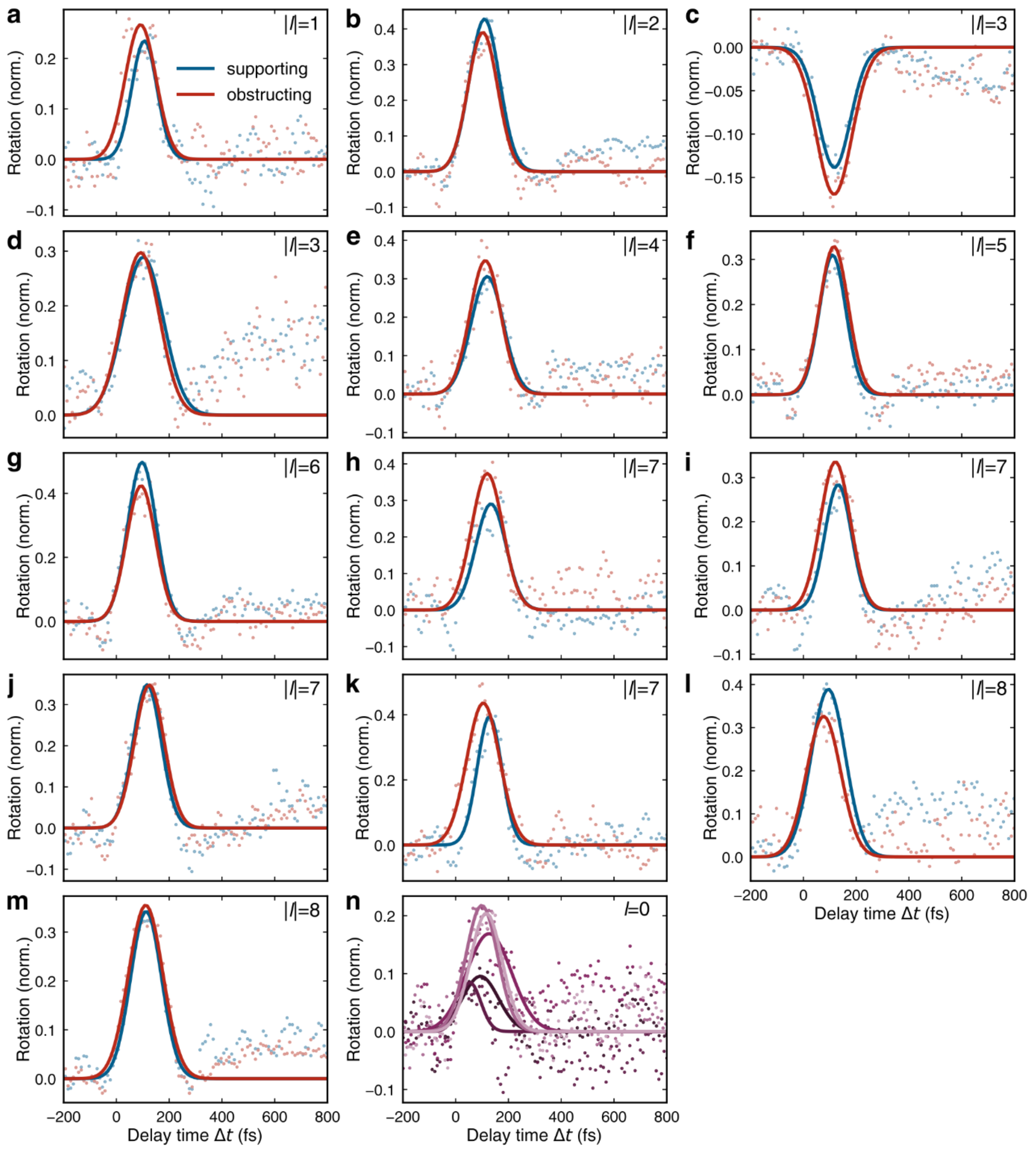

**Supplementary Fig. 8: Fitting of peaks in the nonmagnetic signal contributions of supporting and obstructing excitation geometries for different orders of photonic OAM.** **a**, $|l| = 1$, **b**, $|l| = 2$, **c**, **d** $|l| = 3$ (increasing pump fluences), **e**, $|l| = 4$, **f**, $|l| = 5$, **g** $|l| = 6$, **h-k** $|l| = 7$ (increasing pump fluences), **l-m**, $|l| = 8$ (increasing pump fluences) and $l$=0. The dots represent the experimental data. The blue, red and purple lines correspond to fitted Gaussians for the supporting, obstructing, and $l$=0 cases, respectively.

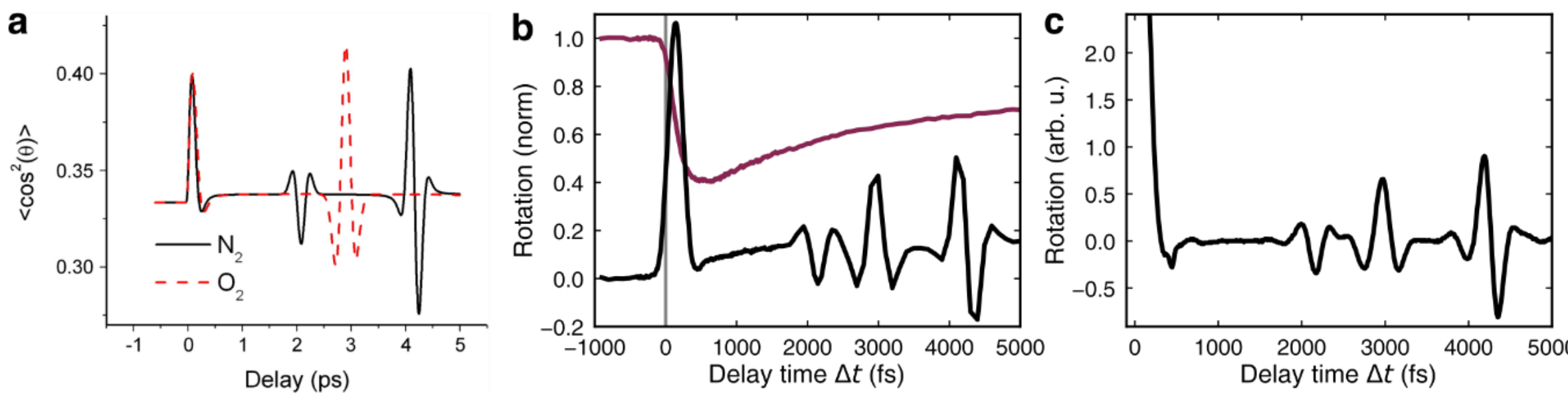

**Supplementary Fig. 9: Pump-excited air affects the probe-pulse polarization rotation. a** Numerical simulations of the alignment factor $<\cos^2(\theta)>$ for $N_2$ (solid black line) and $O_2$ (dashed red line). Laser parameters: 800 nm, 50 fs (FWHM), temporally Gaussian, $I_{peak} = 5 \cdot 10^{13}$ W/cm$^2$. The initial gas temperature is 295 K. **b** Normalized and shifted (in time) magnetic (purple) and nonmagnetic (black) signals from a time-resolved MOKE measurement of 10 nm Ni on MgO. We determined $t_0$ from a fit of the magnetic data and marked its position with the vertical grey line. **c** Polarization rotation of light reflected by a pure MgO sample with no external magnetic field. In this measurement, the pump and probe pulses were linearly polarized with an angle of 45° between them. This results in an additional contribution to the peak near $\Delta t = 0$ fs caused by the electro-optical Kerr effect (EOKE). Note that this data is not normalized. Furthermore, we had no magnetic signal that we could fit to determine the position of $t_0$. The data in **b** and **c** was measured with a different experimental setup than the data in the main manuscript. The excitation geometry is the same, but the laser system used was the model Astrella from Coherent with the same wavelengths but different pulse durations $\tau_{pump} \approx 60$ fs and $\tau_{probe} \approx 50$ fs and a repetition rate of 1 kHz. **a** Reprinted from Supplementary Ref. *1*, Copyright 2010, with permission from Elsevier.

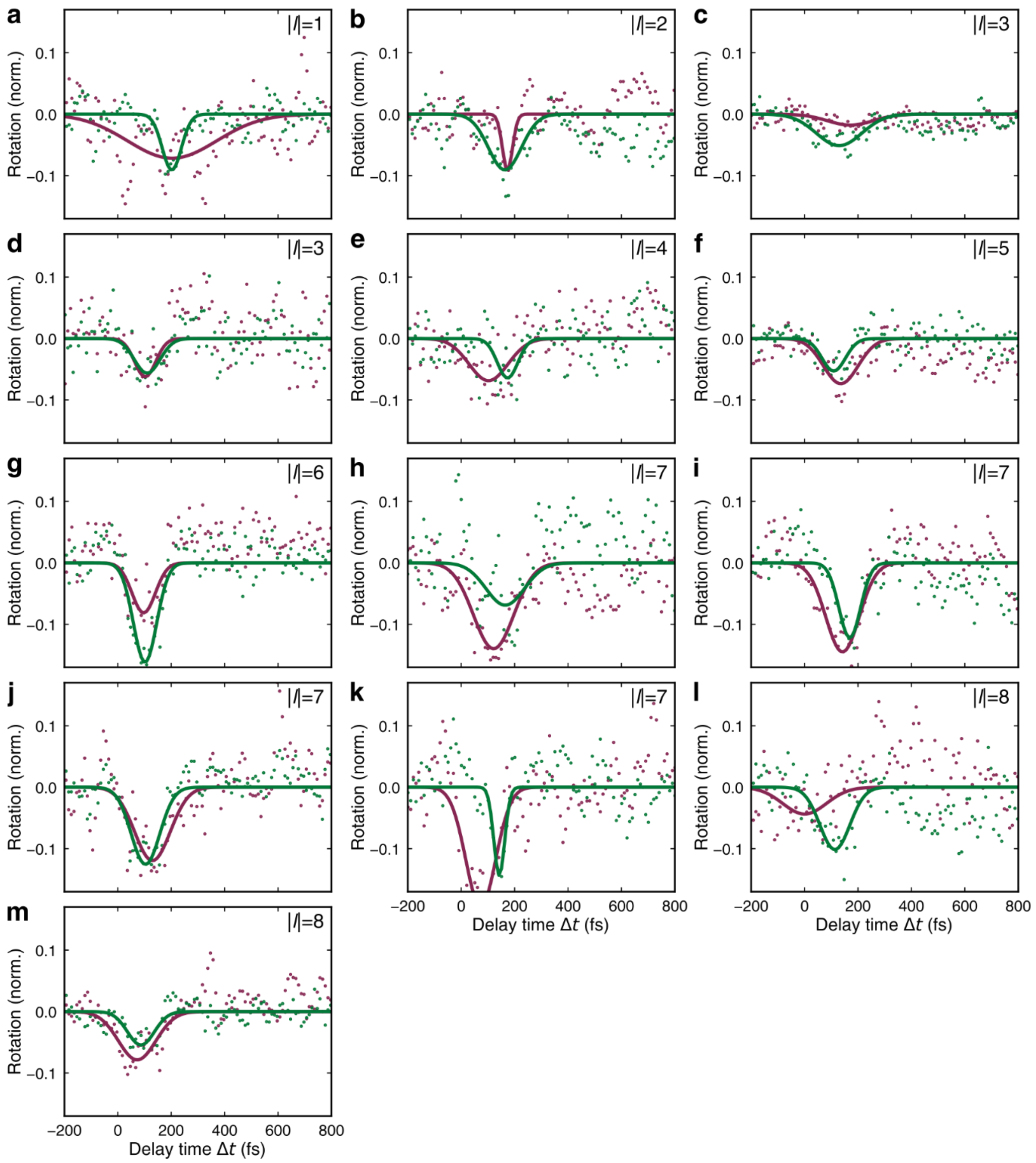

**Supplementary Fig. 10: Fits of the pure OAM contribution to the experimental data with different OAM orders.** **a**, $|l| = 1$, **b**, $|l| = 2$, **c**, **d** $|l| = 3$ (increasing pump fluences), **e**, $|l| = 4$, **f**, $|l| = 5$, **g** $|l| = 6$, **h-k** $|l| = 7$ (increasing pump fluences), and **l-m** $|l| = 8$ (increasing pump fluences). The purple lines and dots correspond to $M^{+}$, and the green lines and dots correspond to $M^{-}$.

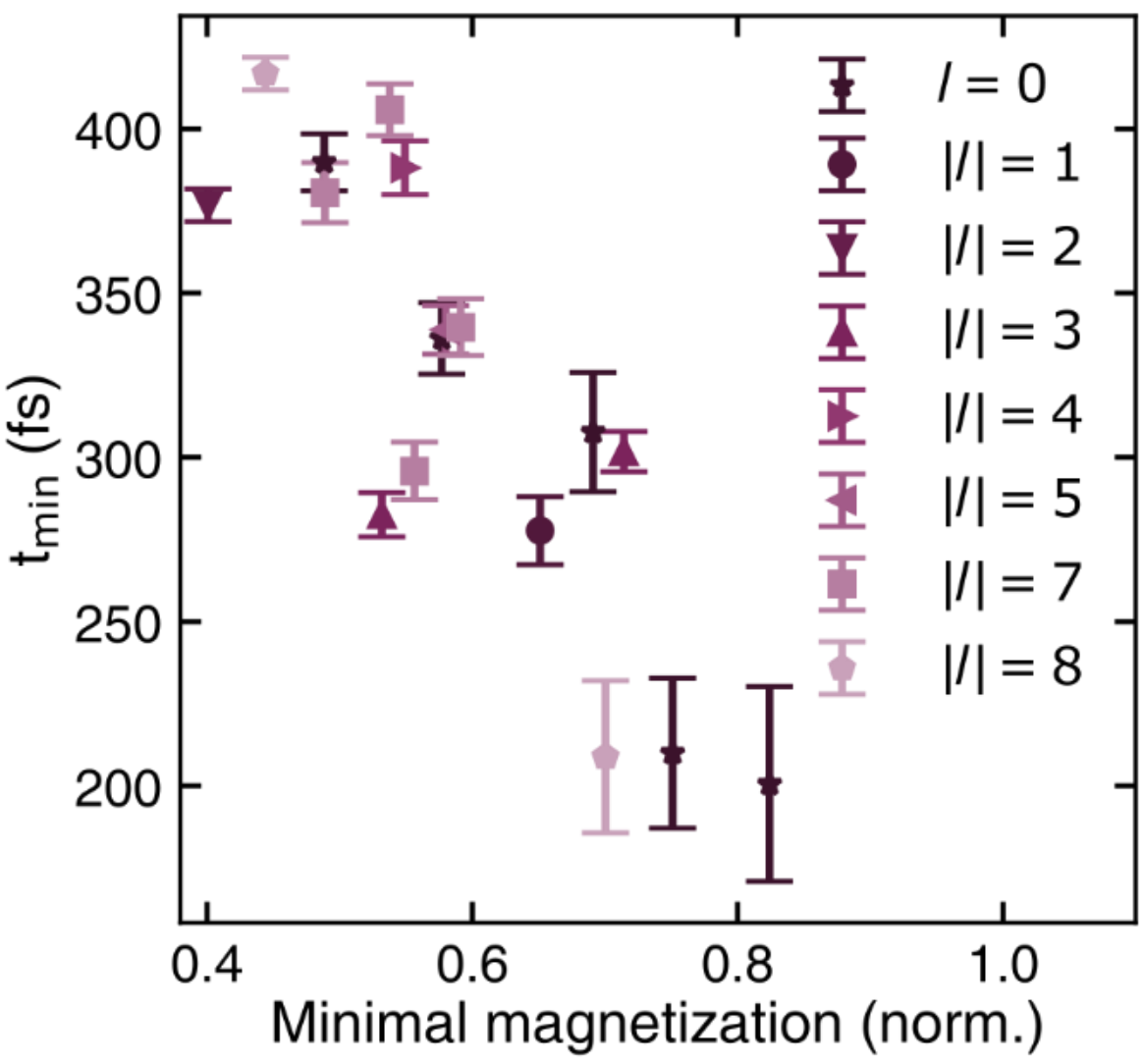

**Supplementary Fig. 11: The OAM order does not influence the demagnetization behavior that is shared by both OAM handednesses (black curves in Fig. 4 and Supplementary Fig. 6 for |*l*| ∈ [1, 8], curves in Supplementary Fig. 5l for *l* = 0).** The delay time $t_{min}$ *is* is plotted vs. the minimal magnetization (minimum of the black curves in Fig. 3 and Supplementary Fig. 7 for |*l*| ∈ [1, 8], this is not $M_{min}$ from equation (1)). The delay time $t_{min}$ represents the time at which the minimum is reached. All OAM orders show a similar behavior; the value of $t_{min}$ increases for a decreasing minimal magnetization (meaning a stronger quenching). This supports the assumption of our model, that the OAM of light adds an additional contribution to the ultrafast demagnetization that can be well approximated with a Gaussian. Because the minimal magnetization value is not a fitting parameter, no error limits are given. The errors of $t_{min}$ are the fitting errors of $t_0$. For $l$ = 6, no data point is plotted because the minimal magnetization was not reached in the measured delay time range.

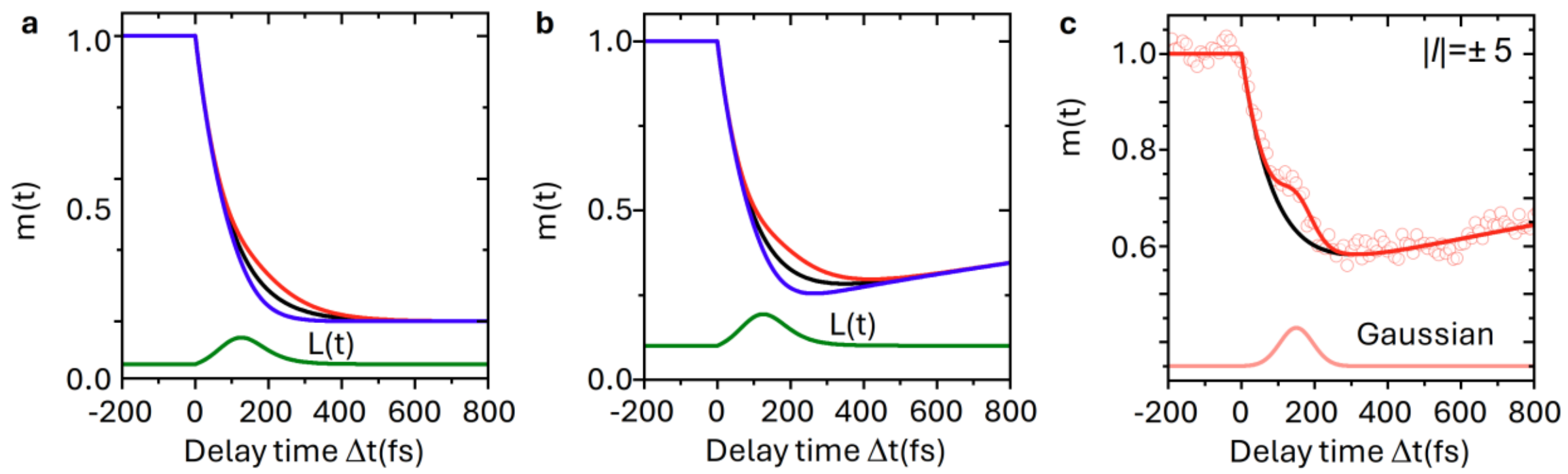

**Supplementary Fig. 12: Temporal evolution of the ultrafast magnetization dynamics predicted by a rate equation model.** The influence of the photonic OAM on the magnetization dynamics was modeled by a rate equation model. The details of this rate equation model are summarized in the *Supplementary Methods* section. **a** Solution of the rate equation model. The predicted time evolution of the conventional magnetization (without photonic OAM) is shown in black, and those for the supporting and obstructing cases are shown in blue and red, respectively. The (angular momentum) population of the photonic OAM-absorbing subsystem is included as a solid green line. The model parameters used to compute these traces are: $\tau_D$= 90 fs, $\tau_T$= 50 fs, $t_A$=90 fs, σ=50 fs, $L_0$=0.15, $M_0$=1. The magnetization traces of the supporting and obstructing cases qualitatively reflect the anomalous line shape of the experimental data. **b** Solution of the rate equation model including an artificial description of the remagnetization dynamics and the final quenching of magnetization. The model parameters used for calculating these traces are: $\tau_D$= 90 fs, $\tau_T$= 50 fs, $t_A$=90 fs, σ=50 fs, $L_0$=0.15, $M_0$=0.8, $\tau_R$= 4000 fs. **c** Fit of the simulated magnetization dynamics for obstructing photonic OAM excitation (red data points) with a linear combination of the common demagnetization dynamics (black) and an additional Gaussian function (light red). The envelope of the fit is shown as a solid red line. The quality of this fit supports the validity of our data analysis procedure using a linear combination of the common demagnetization dynamics and an additional Gaussian curve to model the ultrafast magnetization dynamics after optical excitation with fs light pulses with photonic OAM.

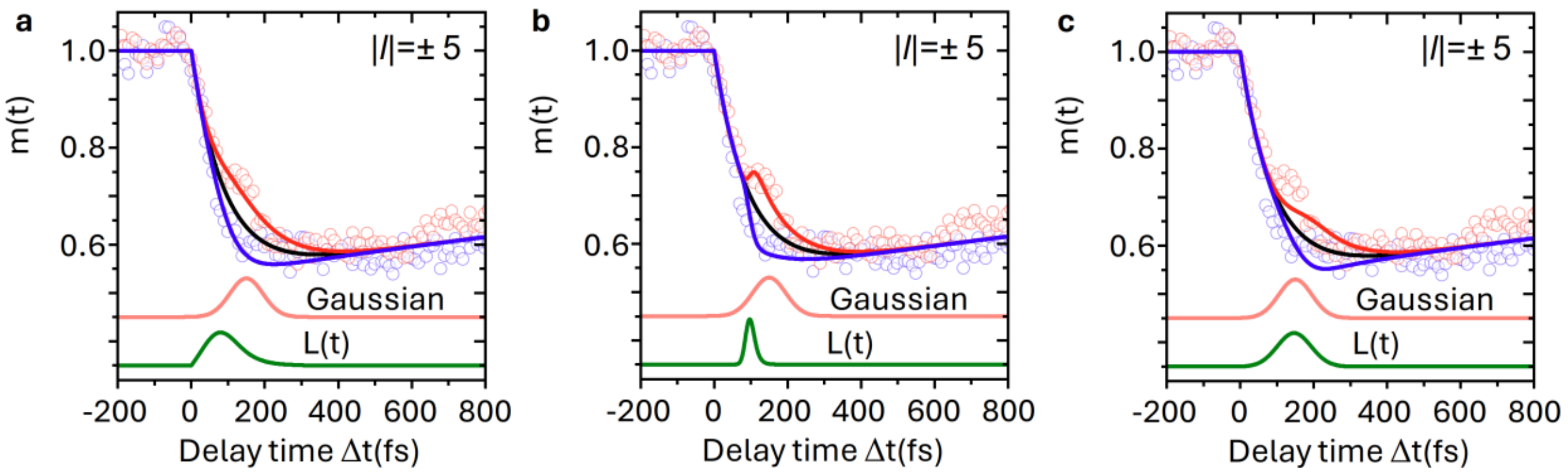

**Supplementary Fig. 13: Comparison of the experimental magnetization dynamics for the supporting and obstructing case for |l|=5 with different solutions of our rate equation model.** The experimental magnetization dynamics for the supporting and obstructing cases (|*l*| = 5) are shown as light blue and red dots. The solutions of our rate equation model are included as solid lines of identical color. The predicted time evolution of the conventional magnetization dynamics (without photonic OAM) is shown in black, the (angular momentum) population of the photonic OAM-absorbing subsystem as a solid green line. The fitted Gaussian-like contribution of the photonic OAM to the magnetization trace is included as lighr red solid line (see Supplementary Fig. 12c for more details). Note that L(t) and the Gaussian curve are shifted vertically for better visibility. The model parameters used for calculating these traces are for **a**: $\tau_D$= 90 fs, $\tau_T$= 50 fs, $t_A$=50 fs, σ=40 fs, $\tau_R$= 4000 fs, $L_0$=0.15, $M_0$=1; **b**: $\tau_D$= 90 fs, $\tau_T$= 10 fs, $t_A$=90 fs, σ=10 fs, $\tau_R$= 4000 fs, $L_0$=0.1, $M_0$=0.47; **c**: $\tau_D$= 90 fs, $\tau_T$= 10 fs, $t_A$=140 fs, σ=45 fs, $\tau_R$= 4000 fs, $L_0$=0.1, $M_0$=0.47. The model parameters used in panels a and b represent the upper and lower bounds of the parameter space that can still reasonably describe the experimental data. Here, we would also like to point out that not all sets of parameters within the upper and lower bounds of the parameter space lead to physically meaningful or likely solutions. For example, for the solution plotted in panel b, the population of L(t) remains zero for several 10 fs before the accumulation of photonic OAM starts. This would indicate a true storage of photonic OAM in another (unknown) degree of freedom before the transfer of photonic OAM into the L subsystem takes place. In addition, we show a solution in panel c with parameters outside the upper and lower bounds of the parameter space. Here, we find a clear deviation of the simulated model curves from the experimental data. In particular, this example shows that the maximum population of L(t) must occur earlier than the maximum effect of the photonic OAM on the magnetization trace. Our estimation of the uncertainty of the model parameters allows us to draw the following conclusions: (i) Our simple rate equation model can describe the anomalous line shape of the OAM-induced magnetization dynamics only in the case of a non-vanishing storage time of the photonic OAM in the material. (ii) There must be a gradual population transfer from the OAM system L(t) to the magnetic system m(t).

## Supplementary Methods

## Rate Equation Model

We modeled the temporal evolution of the anormalous lineshape of the OAM-induced magnetization dynamics by solving a rate equation model for the (angular momentum) population of the magnetic subsystem *m* and a hypothetical subsystem *L*. The subsystem m is responsible for the magnetic order of the material, while the subsystem *L* absorbs and distributes the photonic OAM of the laser pulse in the solid.

In our toy model, we assume that the ultrafast demagnetization due to laser heating (i.e., the intensity of the laser pulse) is fully characterized by the demagnetization time $\tau_D$ (as predicted, for example, by the three-temperature model (Supplementary Ref. *2*)) and is not directly affected by the OAM of the laser pulse. Therefore, we model the ultrafast change of magnetization *m*(*t*) due to laser heating by $\frac{d}{dt}m(t) = -\frac{m(t)}{\tau_D}$.

We further assume that the photonic OAM of the laser pulse is deposited in *L*. The (angular momentum) population of *L* can be transferred to the magnetic subsystem *m* by the unidirectional exchange rate $\frac{1}{\tau_T}$. The source term for the angular momentum in *L* is modeled by a Gaussian function $\frac{L_0}{\sqrt{2\pi}\sigma} \cdot e^{-0.5\frac{(t-t_A)^2}{\sigma^2}}$. $T_A$ is the so-called accumulation time, which accounts for the delayed build-up of OAM inside the solid, e.g. due to an indirect transfer of photonic OAM from the light field into the solid; σ is the temporal width of the OAM pulse in the material. The supporting and obstructing cases are considered by the sign of the amplitude $L_0$. This results in the following set of rate equations:

$$\frac{d}{dt}m(t) = -\frac{m(t)}{\tau_D} + \frac{L(t)}{\tau_T} \quad (1)$$

$$\frac{d}{dt}L(t) = \frac{L_0}{\sqrt{2\pi}\sigma} \cdot e^{-0.5\frac{(t-t_A)^2}{\sigma^2}} - \frac{L(t)}{\tau_T} \quad (2)$$

To solve this set of rate equations, we assume that *m*(*t*) is constant prior to optical excitation ($m(t = 0) = M_0$). Similarly, we require *L*(*t*) to be zero before excitation.

The analytical solutions of the rate equation model are given in equations (3) and (4) and the corresponding m(t) and L(t) curves are plotted in the Supplementary Figure 12(a) for an exemplary set of parameters (see caption for details). Note that we artificially set the value of m(t) to $M_0$=1 for t<0fs, i.e., for the magnetization before the optical excitation.

To compare the solutions of our rate equation model with the experimental data, we modeled the remagnetization dynamics by adding an another (exponential) function $m_R(t)$ to the solution of the rate equation model, see equation (5).

$$m_R(t) = M_0\left(1 - e^{-t/\tau_R}\right) \quad (5)$$

$\tau_R$ is the remagnetization time, and $M_0$ is the magnetization before the optical excitation. In addition, we accounted for the finite quenching of the magnetization in the experimental data by using $M_0<1$ in our simulation. For comparison, we normalized the simulated magnetization trace m(t) to unity before the optical excitation, i.e., m(t<0fs)=1. This procedure changes the lineshape of the simulated magnetization trace for both the supporting and obstructing cases as shown in Supplementary Figure 12b.

$$m(t) = \frac{e^{-t\left(\frac{1}{\tau_D}+\frac{1}{\tau_T}\right)}\left(L_0\tau_D\left(\left(\mathrm{Erf}\left(\frac{\sigma^2+t_A\tau_D}{\sqrt{2}\sigma\tau_D}\right)-\mathrm{Erf}\left(\frac{\sigma^2-t\tau_D+t_A\tau_D}{\sqrt{2}\sigma\tau_D}\right)\right)e^{\frac{t}{\tau_T}+\frac{\sigma^2+2t_A\tau_D}{2\tau_D^2}}+\left(\mathrm{Erf}\left(\frac{\sigma^2-t\tau_T+t_A\tau_T}{\sqrt{2}\sigma\tau_T}\right)-\mathrm{Erf}\left(\frac{\sigma^2+t_A\tau_T}{\sqrt{2}\sigma\tau_T}\right)\right)e^{\frac{t}{\tau_D}+\frac{\sigma^2+2t_A\tau_T}{2\tau_T^2}}\right)-2M_0e^{t/\tau_T}(\tau_T-\tau_D)\right)}{2(\tau_D-\tau_T)} \tag{3}$$

$$L(t) = \frac{1}{2}L_0e^{\frac{\sigma^2-2t\tau_T+2t_A\tau_T}{2\tau_T^2}}\left(\mathrm{Erf}\left(\frac{\sigma^2+t_A\tau_T}{\sqrt{2}\sigma\tau_T}\right)-\mathrm{Erf}\left(\frac{\sigma^2-t\tau_T+t_A\tau_T}{\sqrt{2}\sigma\tau_T}\right)\right) \tag{4}$$

## Supplementary References